\documentclass{article}

\usepackage{arxiv}

\usepackage[utf8]{inputenc} 
\usepackage[T1]{fontenc}    
\usepackage{hyperref}       
\usepackage{url}            
\usepackage{booktabs}       
\usepackage{amsfonts}       
\usepackage{nicefrac}       
\usepackage{microtype}      
\usepackage{lipsum}
\usepackage{graphicx}
\graphicspath{ {./images/} }
\usepackage{multirow}
\usepackage{amsmath}
\usepackage{array}
\usepackage{graphicx}
\usepackage{subcaption}
\usepackage{paralist}

\title{MVIAnalyzer: A Holistic Approach to Analyze \\ Missing Value Imputation\thanks{Software available under \url{https://gitlab.com/d6745/mvianalyzer}}}

\newbox{\orcid}\sbox{\orcid}{\includegraphics[scale=0.06]{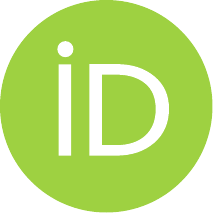}}
\author{
  \href{https://orcid.org/0000-0002-5960-5886}{\usebox{\orcid}\hspace{1mm}Valerie Restat} \\
  University of Hagen \\
  Hagen, Germany\\
  \texttt{valerie.restat@fernuni-hagen.de} \\
  \And
   Kai Tejkl \\
  University of Hagen \\
  Hagen, Germany\\
  \texttt{kai\_te@gmx.net} \\
  \And
  \href{https://orcid.org/0000-0003-2771-142X}
   {\usebox{\orcid}\hspace{1mm}Uta St{\"o}rl} \\
  University of Hagen \\
  Hagen, Germany\\
  \texttt{uta.stoerl@fernuni-hagen.de} \\
}

\begin{document}
\maketitle
\begin{abstract}
Missing values often limit the usage of data analysis or cause falsification of results. Therefore, methods of missing value imputation (MVI) are of great significance. However, in general, there is no universal, fair MVI method for different tasks. This work thus places MVI in the overall context of data analysis. For this purpose, we present the MVIAnalyzer, a generic framework for a holistic analysis of MVI. It considers the overall process up to the application and analysis of machine learning methods. The associated software is provided and can be used by other researchers for their own analyses. To this end, it further includes a missing value simulation with consideration of relevant parameters. The application of the MVIAnalyzer is demonstrated on data with different characteristics. An evaluation of the results shows the possibilities and limitations of different MVI methods. Since MVI is a very complex topic with different influencing variables, this paper additionally illustrates how the analysis can be supported by visualizations.
\end{abstract}

\keywords{missing value imputation \and machine learning \and data cleaning  \and data quality \and analysis}

\section{Introduction}
\label{sec:intro}
In real data, values are often missing. Incomplete data are a potential source of error. For this reason, there are a number of different techniques for replacing missing values, summarized as missing value imputation (MVI).
The importance of MVI is also increasing due to the continuous development of machine learning (ML) procedures, and artificial neural networks as a significant branch of ML.~\cite{Jeong2022} A large number of ML models cannot handle missing data.
Jeong et al. name three points that need to be considered in this context~\cite{Jeong2022}:
\begin{itemize}
    \item A flawed MVI also leads to the adoption of the flaw in the later model.
    \item Due to different MVI methods in training and test, discrimination can occur even with error-free imputation. Details of the original MVI must therefore always be known.
    \item In general, there is no universal, fair MVI method for different tasks. The downstream application must thus already be taken into account in the MVI.
\end{itemize}
Hence, in this paper the MVI is considered in the overall context of data, influences on missing data and the subsequent evaluation of these data. However, MVI cannot be considered in isolation from other issues in the context of data analysis. Both pre- and postprocessing activities must be included. The core objective of this work is thus the consideration of the holistic process, including following steps:
\begin{itemize}
    \item Data import, respectively generation and preparation
    \item Consideration of different characteristics for incomplete data, e.g. missing rate
    \item Application and evaluation of MVI
    \item Application and evaluation of downstream machine learning methods for classification and pattern recognition
\end{itemize}

Consequently, the paper provides the following \textbf{contributions}: It presents the MVIAnalyzer, a generic framework that can be used to holistically investigate MVI. It considers multiple constraints, including different characteristics of incomplete data as well as various MVI methods. Moreover, it provides an independent approach to the analysis with evaluations, to answer the following research questions:
\begin{itemize}
    \item[(1)] How do different constraints and input data influence the quality of MVI?
    \item[(2)] What is the relation between MVI and the results of downstream ML procedures?
    \item[(3)] Can MVI also be neglected in the context of ML procedures?
\end{itemize}
In addition, we provide the software with extensive configuration options, so that it can be used by other researchers for different kinds of studies and objective comparisons in the future\footnote{\url{https://gitlab.com/d6745/mvianalyzer}}. The evaluation carried out is exemplary and does not claim to be complete. Instead, it emphasizes open research questions. The focus is on the provision of the MVIAnalyzer, which provides a basis for further analysis of these research questions. 

The remainder of this paper is as follows: The different characteristics of incomplete data are introduced in Section~\ref{sec:characteristics}. Section~\ref{sec:sota} presents the state of science. Subsequently, our research methodology is described in Section~\ref{sec:research-methodology}. The obtained results are presented in Section~\ref{sec:results}, each for MVI and ML. These parts ends with a conclusion that discusses the research questions presented. As a further step in the analysis, Section~\ref{sec:visual} presents possible visualizations of incomplete data. The final Section~\ref{sec:summary} summarizes the paper.

\section{Characteristics of incomplete data}
\label{sec:characteristics}
The presence of missing values within data sets is subject to certain criteria. These have a significant influence on the suitability and performance of imputation methods and thus also on the results of classifiers based on these data. The basic classification system for describing missing data was introduced by Little and Rubin~\cite{little_rubin} already in 1987. It describes the \emph{missing pattern} as well as the \emph{missing mechanism}.

\begin{figure}[ht]
\centering
  \subfloat[Univariate]{
     \label{fig:kap02:musterfehlenderdaten:univariat}
     \includegraphics[height=4cm]{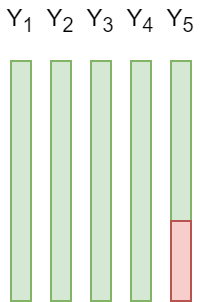}
   } \quad
   \subfloat[Multivariate]{
     \label{fig:kap02:musterfehlenderdaten:multivariat}
     \includegraphics[height=4cm]{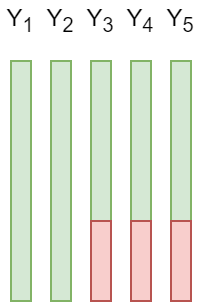}
   } \quad
   \subfloat[Monotone]{
     \label{fig:kap02:musterfehlenderdaten:monoton}
     \includegraphics[height=4cm]{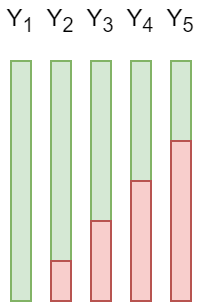}
   } \quad
   \subfloat[General]{
     \label{fig:kap02:musterfehlenderdaten:unbestimmt}
     \includegraphics[height=4cm]{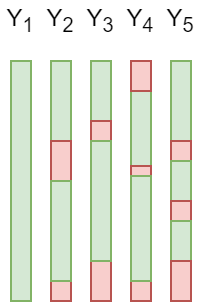}
   } \quad
	\caption{Classification of \emph{missing patterns} according to \cite{little_rubin}, missing values are colored in red (dark for black and white)}
	\label{fig:missing-pattern}
\end{figure}

\paragraph{Missing Pattern}
Figure~\ref{fig:missing-pattern} shows the classification of \emph{missing patterns} according to Little and Rubin~\cite{little_rubin}. Individual attributes are referred to by $Y_i$.

The different patterns are described in more detail below.:
\begin{itemize}
    \item[\textit{Univariate}:] Missing values occur exclusively within a variable, e.g. due to a measurement failure for a certain value.
    \item[\textit{Multivariate}:] Missing values occur in several identical attributes across several data sets, e.g. due to non-response to individual question groups in questionnaires.
    \item[\textit{Monotone}:] Missing values occur consecutively over several attributes in several data sets, e.g. due to increasing failure of responses in long-term studies.
    \item[\textit{General}:] Missing values occur randomly distributed across all data sets in different attributes.
\end{itemize}

\paragraph{Missing Mechanism}
In addition to the missing patterns, the \emph{missing mechanism} according to Little and Rubin~\cite{little_rubin} describes the reason for the absence of values depending on individual attributes:
\begin{itemize}
    \item[\textit{MCAR -- Missing Completely at Random}:] The absence of values of an attribute is completely independent of value expressions of other attributes, e.g. due to the random failure of a sensor.
    \item[\textit{MAR -- Missing at Random}:] The absence of values of an attribute is dependent on the values of another attribute in the same or a previous data collection, e.g. by not answering follow-up questions that do not apply due to previous answers.
    \item[\textit{MNAR -- Missing not at Random}:] The absence of values of an attribute is dependent on the value expression of the same attribute, e.g. due to the removal of certain value ranges from a measurement series by the observer or a measurement value outside the measurement range of a sensor.
\end{itemize}

\begin{figure}[bth]
    \centering
  \subfloat[MCAR]{
     \label{fig:kap02:mcarmarmnar:mcar}
     \includegraphics[height=3.5cm]{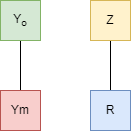}
   } \quad
   \hspace{1cm}
   \subfloat[MAR]{
     \label{fig:kap02:mcarmarmnar:mar}
     \includegraphics[height=3.5cm]{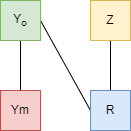}
   } \quad
   \hspace{1cm}
   \subfloat[MNAR]{
     \label{fig:kap02:mcarmarmnar:mnar}
     \includegraphics[height=3.5cm]{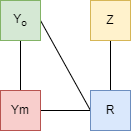}
   }
   	\caption{Graphical representation of missing mechanisms according to~\cite{Schafer2002}, with $Y_o$ as complete attributes, $Y_m$ as incomplete attributes, $R$ as missingness and $Z$ the component of the causes of missingness unrelated to $Y_o$ und $Y_m$.}
 	\label{fig:missing-mechanisms}
\end{figure}

For a better understanding, Figure~\ref{fig:missing-mechanisms} shows a graphical representation of the three missing mechanisms. It should be noted that the respective mechanism does not allow any conclusion to be drawn about the underlying pattern.

MNAR data requires a closer look for the respective reason for failure. Often it is not possible to recover the data and there is no general method for dealing with these missing values~\cite{Garcia-Laencina2010}. The majority of studies are therefore limited to MCAR and MAR mechanisms, and less frequently to MNAR mechanisms (cf.~\cite{Lin2020} and~\cite{Garcia-Laencina2010}).

\paragraph{Missing Rate}
Another influencing factor is \emph{missing rate}. It describes the proportion of data records with incomplete attributes in the total amount of data.

\section{State of science}
\label{sec:sota}
As described, the focus of this paper is on the provision of MVIAnalyzer as an analysis framework. The paper is thus not a survey. This chapter consequently focuses on a general overview. Existing surveys are presented for this purpose, but it explicitly does not claim to cover the most recent papers. The delimitations of this work are further specified at the end of the chapter.

MVI describes different procedures and methods for dealing with and completing incomplete data. As described above, a systematization was already significantly influenced in the 1980s by Little and Rubin~\cite{little_rubin}. However, the number of publications on the use and development of different MVI methods has increased steadily in recent years. This is also shown in the literature review by Adnan et al.~\cite{Adnan2022}. They emphasize the importance of the topic in fields such as medicine, astronomy and computer science and note a significant increase in publications by 2021. 
MVI in connection with artificial neural networks has also received increased attention. Several studies have come to the conclusion that neural networks often achieve better results in MVI than statistical methods or other ML methods~\cite{Shao2017,Yu2021,Che2016}.
MVI can generally be distinguished between \textit{explicit} and \textit{implicit} MVI. In the following, first the state of science of explicit MVI and then implicit MVI is explained. Subsequently, the delimitation of this work is provided.

\subsection{Explicit MVI}
Explicit MVI is associated with a two-step procedure: In the first step, the missing values are imputed by a corresponding MVI method. In the second step, the imputed data is used in different ML methods for classification and pattern recognition.

Various reviews provide an overview of different recent explicit MVI methods~\cite{Adnan2022,Lin2020,Garcia-Laencina2010,Chiu2022,Ren2023}. As in~\cite{Garcia-Laencina2010}, Lin et al.~\cite{Lin2020} first distinguish between statistical and machine learning based MVI methods in their analysis. However, it should be noted that only publications from 2006 to 2017 were considered (111 publications in total), which means that current methods are not listed here.

According to the authors, the most frequently used statistical methods are the expectation-maximization algorithm and the least squares method. Furthermore, simple standard methods such as linear regression or the mean or median method are often used. As the most frequently used machine learning based method, the authors name the KNNI method. It is followed by methods based on decision trees and clustering techniques. It can be seen that due to the limitation until 2017, methods based on neural networks play a minor role compared to later investigations.

In addition to the individual MVI methods, the characteristics of incomplete data are also examined. The authors criticize a frequent limitation of the data. They often contain only a few attributes. Only a small number of studies consider a missing rate >50\%. In about half of all studies, only the MCAR mechanism was considered, less frequently MAR and only in a few individual cases MNAR.

The more recent overview by Chiu et al.~\cite{Chiu2022} also shows similar results. Three quarters of the papers contain only missing rates below 50\%. Although MAR was most frequently identified as a missing mechanism, MNAR plays a minor role as well. The studies are based on 48 publications between 2011 and 2021. Due to the greater recency, an increase in methods based on neural networks can be seen.

One of the most comprehensive reviews came from Adnan et al.~\cite{Adnan2022}. They examined a total of 430 publications from 1991 to 2021. However, the focus of this work was on the metadata (authors, keywords, citation, etc.). No content analysis was carried out. For this reason, no information is available on the individual MVI methods. 

Ren et al.\cite{Ren2023} also conducted a comprehensive review of the challenges that arise in case of missing data. For this, they analyzed 63 publications and compared different MVI strategies. In addition, they conducted experiments with typical MVI methods on different data sets. The results show that most publications focus on MAR and only a few consider all three missing mechanisms. Their review findings also show that the analysis of missing patterns is an open research challenge.

According to Samad et al.~\cite{Samad2022}, one of the most widely used MVI methods is currently the \textit{MICE} method. In studies, it showed a significantly more robust classification performance in comparison to KNNI, depending on the missing rate. Both methods were clearly superior to the median method.

\subsection{Implicit MVI}
In contrast to the two-step procedure of the explicit MVI, the implicit MVI takes the missing values directly into account in the actual model of the application. However, there are far fewer studies on this. The main reason is probably that many ML methods cannot handle \textit{null} values\cite{Ennett2011}. One exception to this are decision trees, for which there are different methods for taking missing values into account. Examples are the \textit{FAIR MIP Forest}~\cite{Jeong2022} or the \textit{MIA procedure} after~\cite{Twala2008}. Another possibility, according to Garcia-Laencina et al.~\cite{Garcia-Laencina2010}, is ensemble methods from neural networks (e.g.~\cite{Krause2003,Jiang2005}). However, these have some disadvantages. For example, some methods cannot use all available data for training. Furthermore, the training time increases.

Another alternative is the \textit{artificial neural network missing inputs dropout (ANN-MIND)} method according to~\cite{Mudau2018}. This is based on the \textit{neural network dropout} according to~\cite{Srivastava2015}. According to the authors, this method showed better performance on different data compared to other methods~\cite{Mudau2018}.

\subsection{Delimitation of this work}
A large number of different methods for MVI already exist. In some cases, broadly scattered data of different sizes are also used to prove the generalization capability. In most cases, however, the following limitations are present:
\begin{itemize}
    \item Missing rate under 50\%
    \item Few missing pattern
    \item In most cases, only MCAR is considered as missing mechanism
\end{itemize}

In contrast, this paper considers a general and extensible approach. This provides a basis for future analyses based on common procedures. The focus is not on a direct comparison of individual MVI methods, but on a holistic view of the entire process. The following points are of particular interest:
\begin{itemize}
    \item Generic framework for the comparison of MVI methods
    \item In addition to the evaluation of the MVI methods, their effect on the performance of downstream ML methods is also investigated
    \item The characteristics of incomplete data is taken into account and systematized as comprehensively as possible by a missing value simulation
\end{itemize}

\section{Research methodology}
\label{sec:research-methodology}
In the following, we will describe the methodology for analyzing the quality and impact of MVI methods on different ML methods. The process is initially limited to tabular data. However, an extension to e.g.\ image data is possible.

\subsection{Procedure}
\label{subsec:procedure}
\begin{figure*}[ht]
  \centering
  \includegraphics[width=\linewidth]{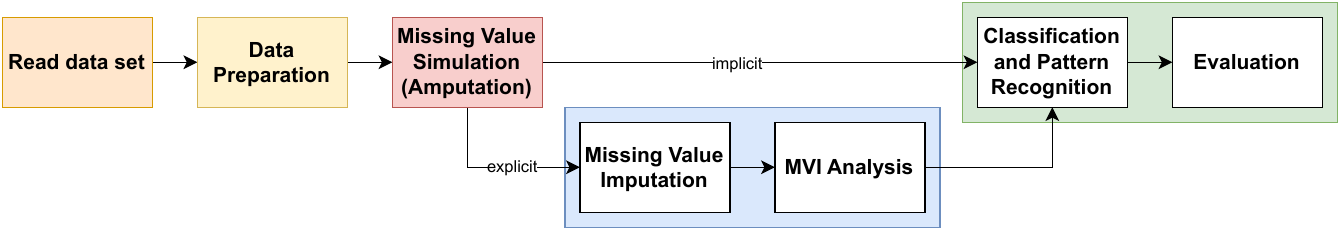}
  \caption{Five steps for analyzing the quality and impact of MVI methods on different ML methods}
  \label{fig:process}
\end{figure*}

The general process of the research methodology can be divided into five steps. These are shown in Figure~\ref{fig:process} and are described in more detail in the following sections.

\subsubsection{Data Preparation}
\begin{figure*}[ht]
\centering
  \subfloat[Original values]{
     \label{fig:kap04:personenpseudodaten_sample_original_prepared:original}
     \includegraphics[height=3.5cm]{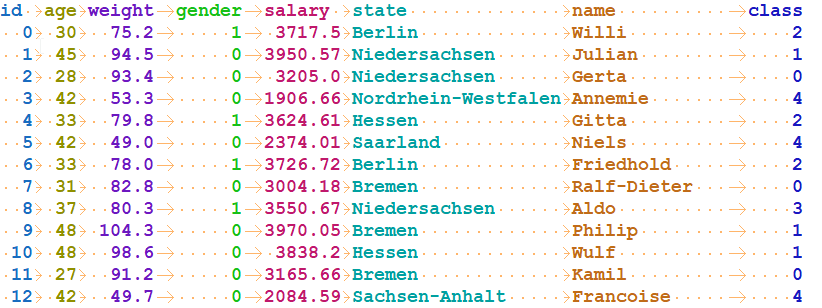}
   } \quad
   \hspace{0.3cm}
   \subfloat[Preprocessed values]{
    \label{fig:kap04:personenpseudodaten_sample_original_prepared:prepared}
     \includegraphics[height=3.5cm]{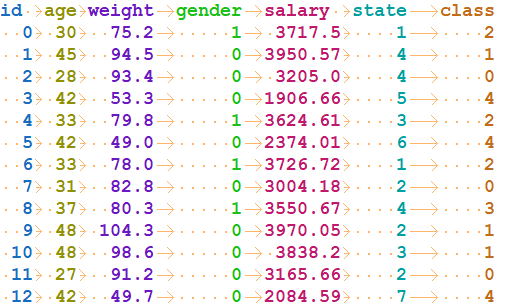}
   } 
   \caption{Comparison of (a) original values and (b) preprocessed values}
   \label{fig:sample_original_prepared}
\end{figure*}

After reading the data set data preparation includes \textit{encoding}, \textit{normalizing} and \textit{splitting} the data. 
Figure~\ref{fig:sample_original_prepared} shows an extract of a data set with the original and preprocessed values in comparison.

For the methods used, all data had to be in numerical form. All non-numeric attributes classified as categorical were therefore converted in advance. For a consideration of non-categorical attributes (such as \textit{name}), extended models are necessary, which were not considered in the context of this work.

\begin{figure}[ht]
	\begin{center}
		\includegraphics[width=0.5\linewidth]{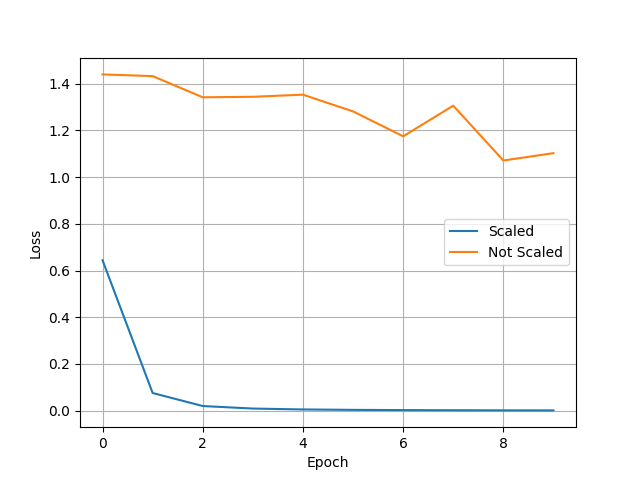}
		\caption{Comparison of loss values with and without normalization}
		\label{fig:scaling}
	\end{center}
\end{figure}

To avoid attributes with a larger value range being weighted more heavily when using neural networks, all attributes must first be normalized to the same value range (see also~\cite{Khan22,Yoon18}).
Figure~\ref{fig:scaling} shows an example of the effect of normalization for a classification with a standard multi-layer perceptron\footnote{Three hidden layers with 265/128/64 neurons} for one of the data sets used. The loss values during a test run over 10 epochs are shown. This underlines the need for normalization.
Normalization was not carried out as part of the preprocessing, but only during the actual model application for MVI or classification and pattern recognition. This ensures that the original values are available throughout the evaluation process. The normalized values are only determined at the necessary time and transformed back after use. This simplifies the evaluation of the individual steps in the comparison with the original data.

Subsequently, the data was divided as standard into 80\% training data, 15\% test data and and 5\% validation data.

\subsubsection{Missing Value Simulation (Amputation)}
One of the objectives of this work, as described, is to investigate the impact of different characteristics of incomplete data on machine learning analysis. In this respect, the broadest possible simulation of missing values, taking into account the different influencing factors, represents a decisive step for the subsequent evaluation of the models.

For this reason, a missing value simulation (MVS) was carried out. Here, the amputation procedure according to Schouten et al.~\cite{Schouten2018} and its implementation in the package \textit{pyampute}~\cite{Schouten_pyampute_2022} was used.

The different characteristics presented in Section~\ref{sec:characteristics} were taken into account. The large number of different manifestations results in a large number of different combinations. For this reason, restrictions had to be made. In order to cover as many relevant cases as possible, the following variants were chosen: For each of the~\textbf{4} missing patterns (\textit{univariate}, \textit{multivariate}, \textit{monotone}, \textit{general}), the MVS was performed with all \textbf{3} missing mechanisms (\textit{MCAR}, \textit{MAR}, \textit{MNAR}). In addition, \textbf{5} different missing rates (\textit{0.1}, \textit{0.3}, \textit{0.5}, \textit{0.7}, \textit{0.9}) were considered for each possible combination. In total, this resulted in $4 * 3 * 5 = 60$ examination cases for the MVS. The respective target attribute for later machine learning models is excluded from the amputation. It therefore always remains as a complete attribute for later comparison.

\subsubsection{Imputation}
For the explicit MVI, different imputation methods were applied to the data generated by the MVS in the previous step. Both statistical methods and ANN-based methods were used.

In the area of statistical methods, the mean and median imputation was chosen as the baseline. KNNI according to~\cite{Troyanskaya2001} and the MICE method were also considered. With KNNI, the missing values are determined on the basis of the mean value of the k-nearest neighbours. In contrast to KNNI, the MICE method is a multivariate imputation. Each incomplete attribute is estimated via a function of the other attributes. 

In addition to statistical methods, ANN-based methods were also applied for MVI. The implementation was oriented to the procedure described in~\cite{Garcia-Laencina2010}:
\begin{compactenum}
    \item All data sets $X$ are first divided into complete data sets $X_o$ and incomplete data sets $X_m$ ($o = observed$, $m = missing$).
    \item For each missing pattern in $X_m$, a model is created based on $X_o$. This model uses the incomplete attributes of $X_m$ as target attributes. One model is therefore required for each missing pattern.
\end{compactenum}

Depending on the missing pattern, two different types of models are required:
\begin{compactenum}
    \item If the missing pattern only has an incomplete attribute $X_m$ and $X_m$ is a categorical attribute, a classifier can be used as a model.
    \item For all other missing patterns, a regressor model is used.
\end{compactenum}

\subsubsection{MVI Analysis}
In the first step of the analysis, the methods of explicit MVI are evaluated. The original values are compared with the imputed values. For each attribute, the evaluation criteria are determined. Categorical attributes are evaluated by means of accuracy (ACC). For non-categorical attributes, the evaluation is done using the NRMSE (Normalized Root Mean Squared Error). For each characteristic, the results were summarized. The evaluation is based on the mean value of the evaluation criterion. For this purpose, the results were grouped according to missing pattern, missing rate, missing mechanism, and MVI method and the mean value of the evaluation criterion was calculated for the remaining test cases.

To determine the evaluation criteria, only the imputed data sets were considered in relation to the corresponding values of the original data. Thus, a bias of the results is avoided, since the proportion of identical values in the imputed data and the original data directly depends on the respective missing rate. An adjustment had to be made, though, for the calculation of the NRMSE. The minimum and maximum values of the original data are not determined on the above-mentioned reduced original data set but on the entire original data set. Exclusive restriction to the range of values of the original data in the imputed data sets does not allow correct calculation of the NRMSE. Imputated values could also fall outside of this range. For this reason, the term NRMSE2 is used in the following, which is given as:
\begin{equation} \label{eq:nrmse2_minmax}
	NRMSE2_{k} = \frac{RMSE_k}{Max(X_{ges,k}) - Min(X_{ges,k})}
\end{equation}

\subsubsection{Classification and Pattern Recognition}
In the second step of the analysis, different ANN-based models are applied. For the explicit methods, the imputed data are used. The following methods are applied as implicit techniques:
\begin{compactitem}
    \item \textit{delete}: Incomplete data is removed. Classification and pattern recognition is done only on the remaining complete data.
    \item \textit{zerofill}: Incomplete data are replaced by 0-values. The classification and pattern recognition is then carried out on the total data, as in the explicit method. However, since no explicit MVI method is used, this is counted as an implicit method. 
\end{compactitem}
Due to a lack of implementations, the ensemble methods presented in Section~\ref{sec:sota} as well as the ANN-MIND could not be examined in more detail. This has to be carried out in future work.

The ANN-based ML methods applied within this work are limited to the classification of the data with respect to a target attribute. This was previously defined in the respective data in form of a categorical attribute.
Depending on the number of neurons on the input layer as well as on the output layer, an approach with consideration of different topologies was chosen for classification:
\begin{itemize}
    \item ANNs with two, three and four layers were considered.
    \item For the number of neurons in the hidden layers, three cases were considered:
    \begin{itemize}
        \item Minimum: The sum of all neurons in all layers is equal to the size of the input layer.
        \item Medium: The sum of all neurons in all layers is equal to the size of the input layer plus output layer.
        \item Maximum: The sum of all neurons in all layers is equal to twice the size of the input layer plus output layer.
    \end{itemize}
    \item The number of neurons per layer always decreases in the direction of the output layer.
\end{itemize}

In addition to ML models for classification, self-organizing maps (SOM)~\cite{Esaki2017} were used in this work. These are methods of unsupervised learning. SOM models are usually two-layer networks with one input and one output layer. Accordingly, the SOM topology has less complexity. The size of the input layer corresponds to the dimension of the input data, thus the number of attributes.

An important parameter is the number of neurons in the output layer. These directly determine the edge length of the resulting topographic map. However, there are no clear specifications for this. A reference value according to~\cite{Tian2014} was therefore used. Following this, the number of neurons of the output layer $n_{neurons}$ can be estimated as a function of the number of available data sets $N$ as follows:
\begin{equation}  \label{eq:som_neurons}
	n_{neurons} \approx 5 * \sqrt{N}
\end{equation}
Within this work only two-dimensional maps are considered. The edge length $L$ thus results as follows:
\begin{equation}\label{eq:som-L}
	L = \sqrt{n_{neurons}} = \sqrt{5 * \sqrt{N}}
\end{equation}
The edge length L is rounded down to the nearest integer value. In order to be able to additionally estimate the influence of the edge length, three different topologies of SOM models are considered within each of the individual investigations:
\begin{itemize}
    \item Minimum: $L_{min} = 0,25 * L$
    \item Medium: $L$ according to equation~\ref{eq:som-L}
    \item Maximum: $L_{max} = 4 * L$
\end{itemize}

For the evaluation of the ML models, the F1 makro value was applied. 
\begin{equation}
	F_{{1,makro}_{k}} = \frac{1}{J} \sum_{j=1}^{J} F_{1_{k,j}}
\end{equation}
This is suitable for unequally distributed class memberships of an attribute. Based on the sensitivity and precision, the F1 measure is determined for each class $j=1 ..J$ for an attribute $k$ and the mean value is taken over all classes.

\subsection{MVIAnalyzer}
All analyses were carried out using a specifically implemented MVIAnalyzer. This was developed particularly against the background of mapping the entire process -- from data import and generation, MVS, MVI and the machine learning application. Execution is based on the process presented in Figure~\ref{fig:process} and described in Section~\ref{sec:research-methodology}. The generic approach is implemented by means of a class-based, object-oriented implementation with a corresponding flow script. The calculation process can be carried out specifically in a few steps. Re-use of individual classes is possible. Moreover, the integration of further MVI and ML procedures is possible by the additional supply of appropriate classes. The generalization of the class for data loading also enables the integration of further data formats by implementing the corresponding interfaces. The whole execution can be adapted to different researches by configuration files. These files are created in YAML format. The implementation is done in Python (version 3.10.8). Additional, freely available packages were used. Thus, it can be applied and extended universally. Via a Dockerfile, the application can also be used in isolation in a virtual container environment. In addition, the deployment of a Jupyter Notebook allows the execution also within the Google Colab environment.
The corresponding framework is provided in Gitlab\footnote{\url{https://gitlab.com/d6745/mvianalyzer}}. The respective data model and corresponding description can also be found there.

\subsection{Data}
The results presented in this paper were obtained based on synthetic data called \textit{PersonPseudo Data}. Synthetic data offer the advantage of easy adaptation to the context. For example, the number of data sets and the attributes and their distribution can be chosen arbitrarily. We also conducted initial tests with real data. These showed similar results as with the synthetic data. In the future, we want to carry out more analyses with real data. In the context of this work, only tabular data was examined. Nevertheless, the framework developed can in principle also be used for image data, as these can also be represented numerically by specifying the color values of individual pixels.  

\begin{figure}[ht]
	\begin{center}
		\includegraphics[width=0.6\linewidth]{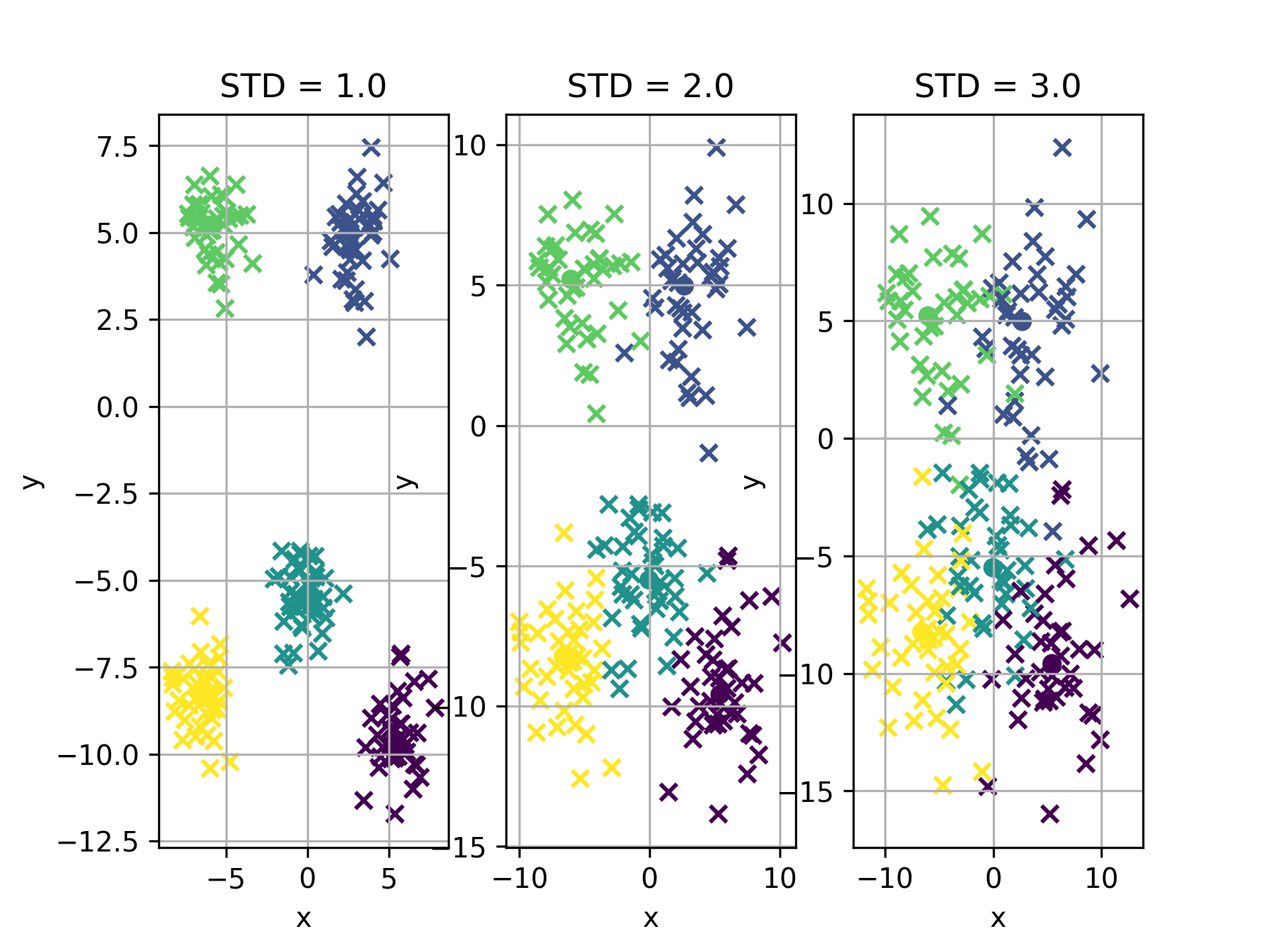}
		\caption{Example of the generation of random x-y value pairs for different standard deviations. The coloring of the individual points indicates the respective cluster membership.}
		\label{fig:make_blobs_sample}
	\end{center}
\end{figure}

Within the MVIAnalyzer, clustered random Gaussian distributed data was generated using a pseudo-data generator. In addition to the range of values, the distribution of the data -- and thus the dispersion within a cluster -- can be defined with the standard deviation. Figure~\ref{fig:make_blobs_sample} shows an example with 200 values and two attributes ($x$ and $y$) for five clusters each at different standard deviations.

\begin{figure}[ht]
	\begin{center}
		\includegraphics[width=0.7\linewidth]{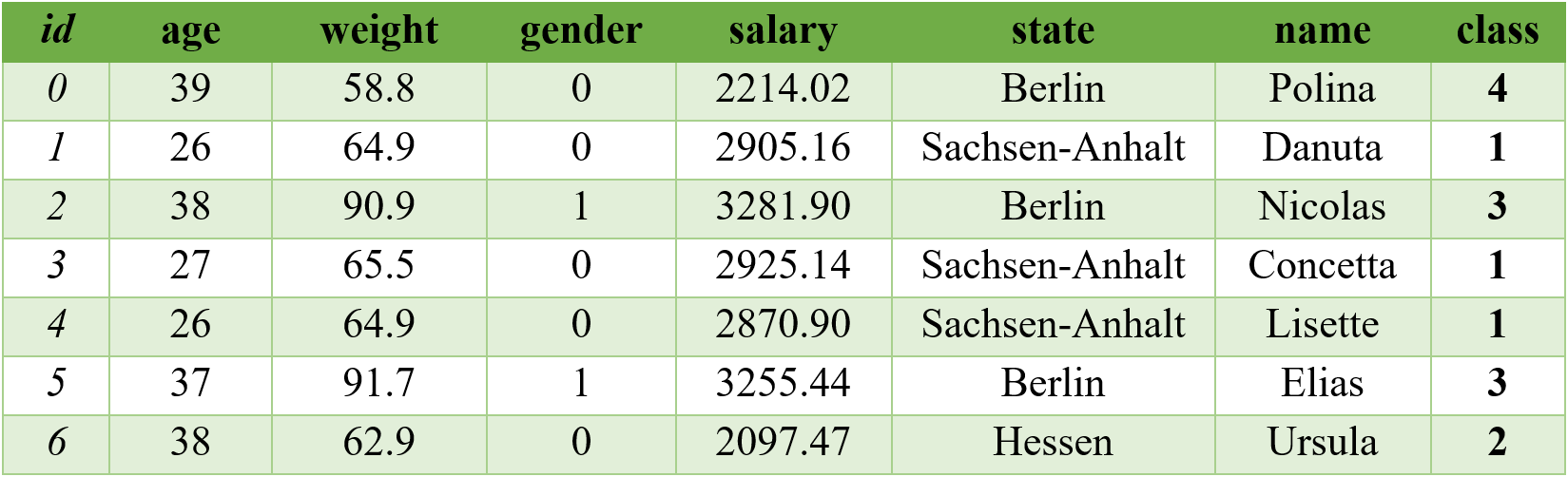}
		\caption{Extract from \emph{PersonPseudo Data}}
		\label{fig:person_pseudo_data}
	\end{center}
\end{figure}

In this way, PersonPseudo Data were created for this work. In order to establish a certain reference to reality for the further evaluations, individual attribute values were manually adjusted. In total, the data set consists of seven attributes, including a \textit{class} attribute that denotes the respective cluster. Figure~\ref{fig:person_pseudo_data} shows an excerpt from seven randomly generated data sets.
These data were subjected to the preprocessing steps and the MVS as well as MVI described in Section~\ref{subsec:procedure}.

\section{Results}
\label{sec:results}
This section presents the results of our analyses. First, the MVI is considered. The following research question from Section~\ref{sec:intro} was examined: \textit{(1) How do different constraints and input data influence the quality of MVI?} The machine learning part is considered subsequently. The objective was to answer the following research questions: \textit{(2) What is the relation between MVI and the results of downstream machine learning procedures?} \textit{(3) Can MVI also be neglected in the context of machine learning procedures?} Both sections are closed with a conclusion in which the corresponding research questions are discussed.

\subsection{Missing Value Imputation}
In the following, the results regarding missing pattern, missing rate, missing mechanism, and MVI methods are shown. It should be noted that for
NRMSE2 smaller values, whereas for ACC larger values represent a better result.

\begin{figure}[ht]
	\begin{center}
		\includegraphics[width=0.9\linewidth]{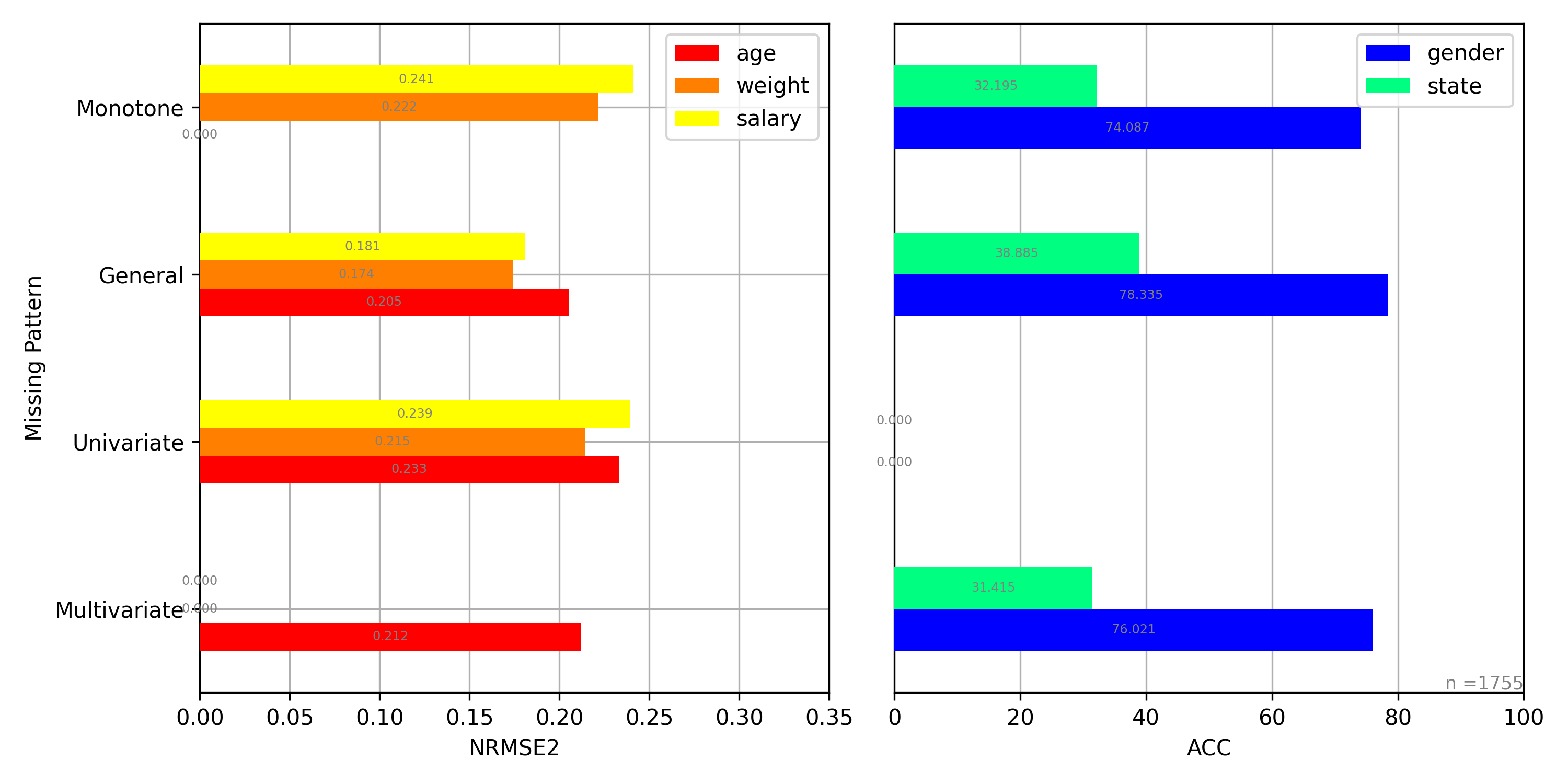}
		\caption{Average NRMSE2 / ACC values over all cases of the PersonPseudo Data depending on \textit{missing pattern}}
		\label{fig:ppd_standard_bar_amppattern}
	\end{center}
\end{figure}

For the \textit{missing pattern}, the number of simultaneously amputated attributes can be identified as the main influencing factor. Univariate and general  missing patterns showed better results than the other missing patterns, as can be seen in Figure~\ref{fig:ppd_standard_bar_amppattern}.

\begin{figure}[ht]
	\begin{center}
		\includegraphics[width=0.9\linewidth]{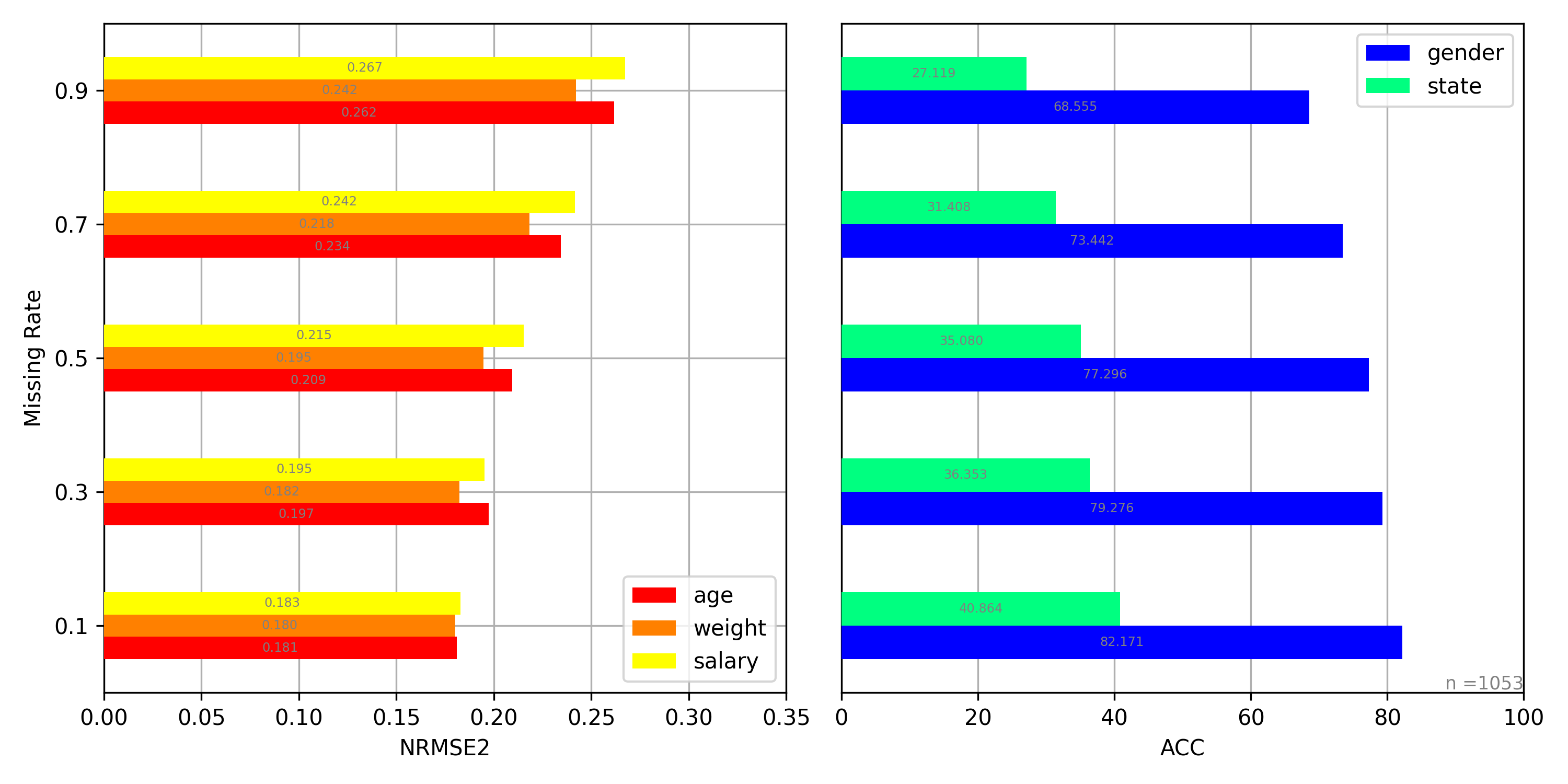}
		\caption{Average NRMSE2 / ACC values over all cases of the PersonPseudo Data depending on \textit{missing rate}}
		\label{fig:ppd-missing-rate}
	\end{center}
\end{figure}

The results on \textit{missing rate} are shown in Figure~\ref{fig:ppd-missing-rate}. As expected, there is a trend towards deterioration of the results with increasing missing rate.

\begin{figure}[ht]
	\begin{center}
		\includegraphics[width=0.9\linewidth]{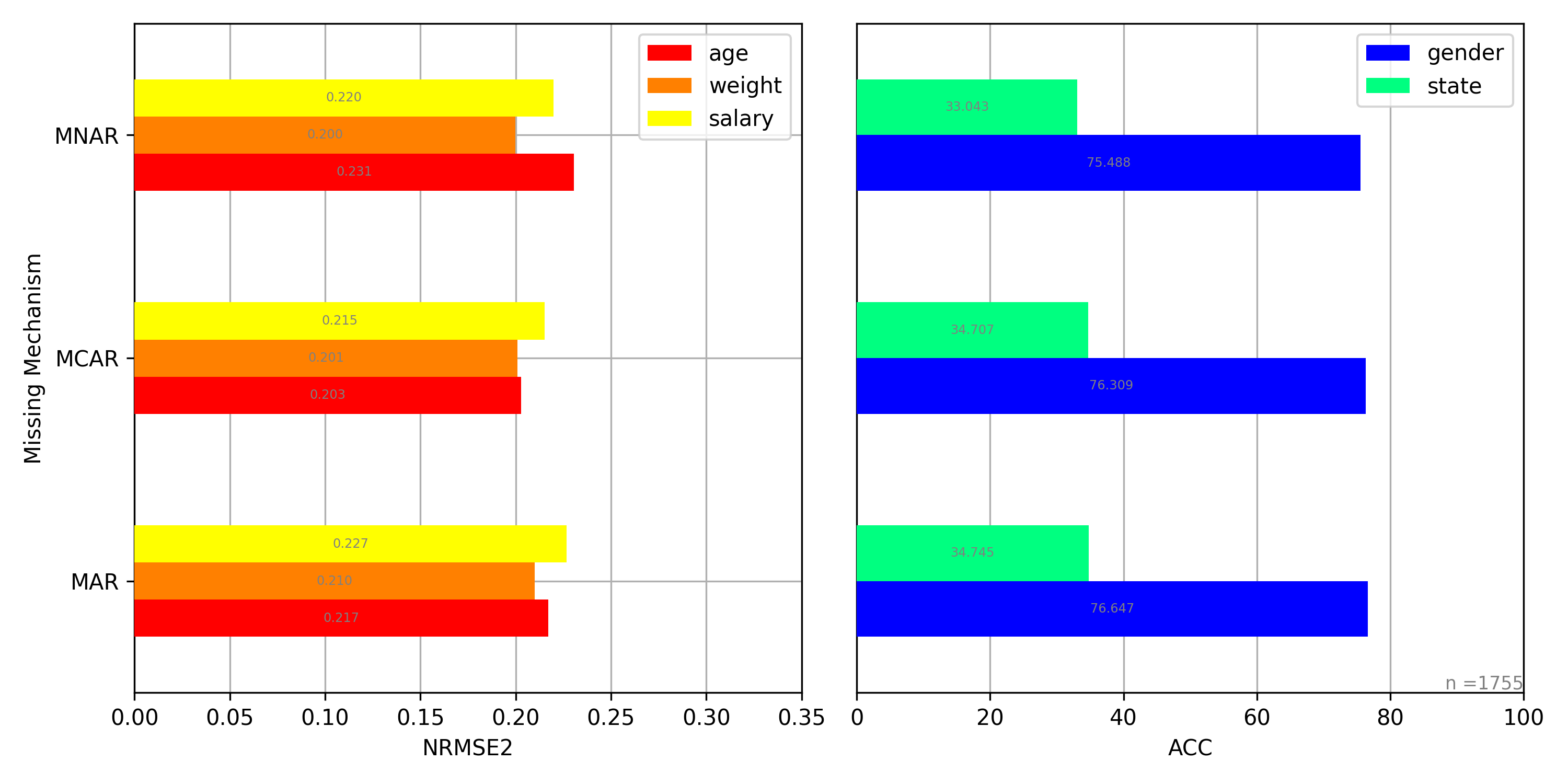}
		\caption{Average NRMSE2 / ACC values over all cases of the PersonPseudo Data depending on \textit{missing mechanism}}
		\label{fig:ppd-missing-mechanism}
	\end{center}
\end{figure}

The results depending on the \textit{missing mechanism} are shown in Figure~\ref{fig:ppd-missing-mechanism}. MNAR achieves slightly worse results. The best results were found for MCAR. However, the differences between the individual mechanisms are only minor.

\begin{figure}[ht]
	\begin{center}
		\includegraphics[width=0.9\linewidth]{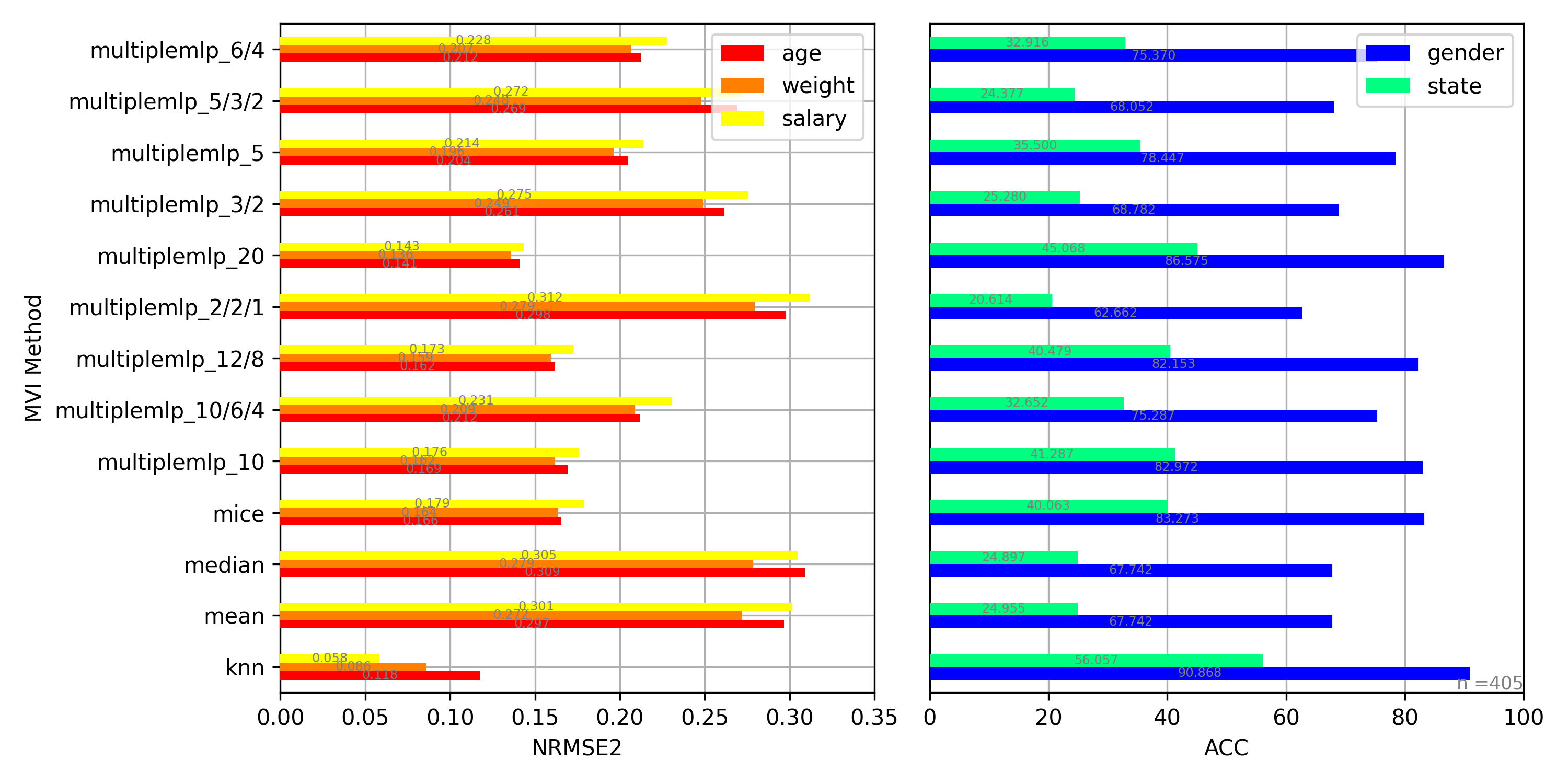}
		\caption{Average NRMSE2 / ACC values over all cases of the PersonPseudo Data depending on the \textit{MVI method}}
		\label{fig:ppd-mvi-methods}
	\end{center}
\end{figure}

Different methods were used for the MVI. The results are presented in Figure~\ref{fig:ppd-mvi-methods}. 
The neural networks are referred to here as \textit{multiplemlp<\_neurons\-\_per\_layer>}. The best results were obtained by imputation with KNNI. This is mainly due to the cluster structure of the synthetic data. As expected, mean and median methods show significantly worse results. The MICE method also achieved good results. For the MLPs, the results depend strongly on the number of hidden layers. With a decreasing number of hidden layers, a clear deterioration of the results can be seen.

\begin{figure}[ht]
	\begin{center}
		\includegraphics[width=0.9\linewidth]{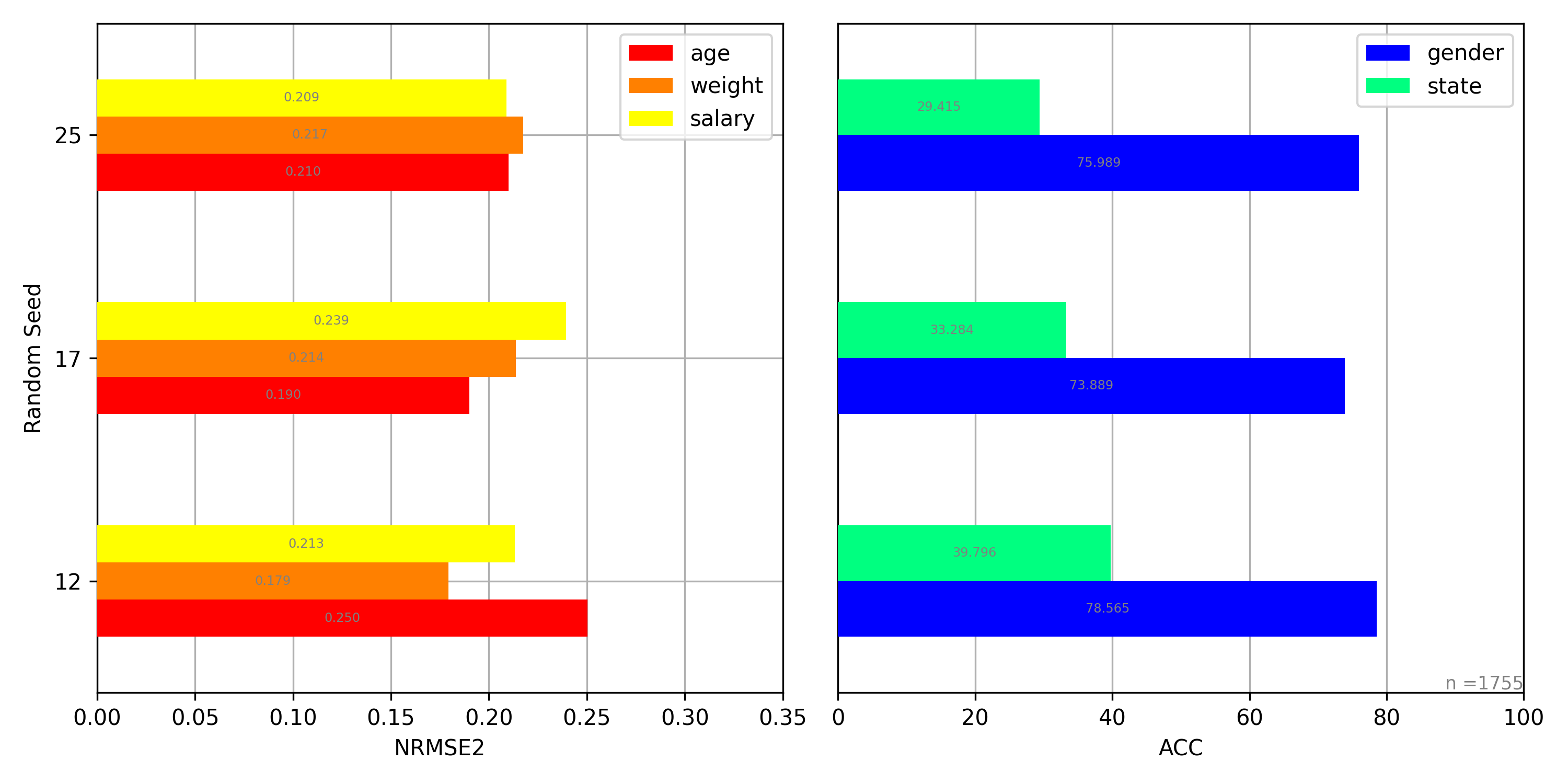}
		\caption{Average NRMSE2 / ACC values over all cases of the PersonPseudo Data depending on the \textit{random seed} for data generation}
		\label{fig:ppd-random-seed}
	\end{center}
\end{figure}

In the case of synthetic data, the influence of different data generation parameters can also be analyzed. In the case of the PersonPseudo data, the random seed and the dispersion of the data were examined. Figure~\ref{fig:ppd-random-seed} shows that different seeds have no significant influence on the quality of the MVI. The previously considered results are therefore independent of the individual values.

\begin{figure}[ht]
	\begin{center}
		\includegraphics[width=0.9\linewidth]{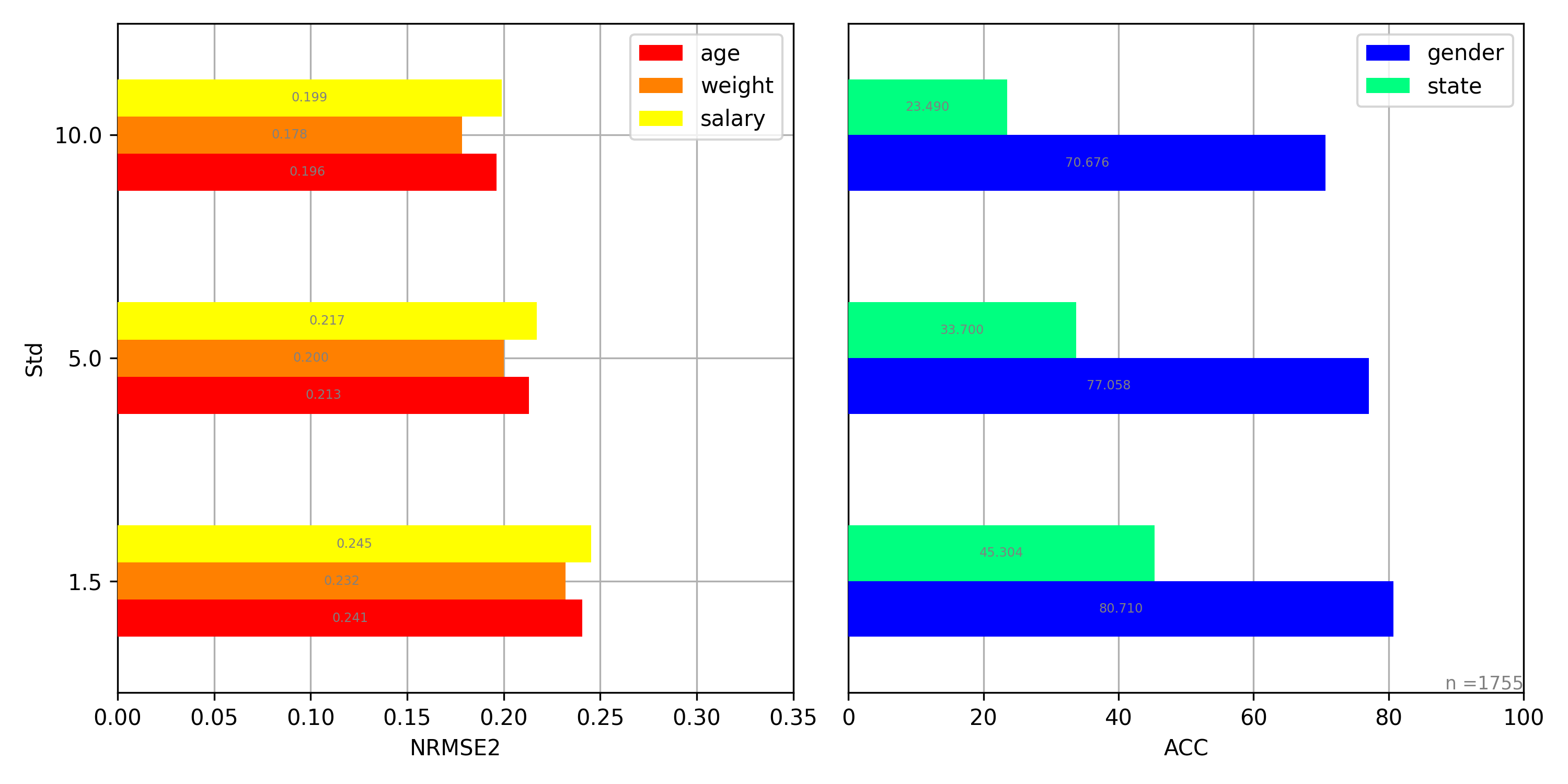}
		\caption{Average NRMSE2 / ACC values over all cases of the \emph{PersonPseudo Data} depending on the standard deviation for data generation}
		\label{fig:ppd-std}
	\end{center}
\end{figure}

The dispersion of the data, on the other hand, has a greater influence, as can be seen in Figure~\ref{fig:ppd-std}. The NRMSE value improves with a higher standard deviation. On the other hand, the accuracy deteriorates. This can be explained by the fact that data with a higher standard deviation have a larger range. This leads to a decrease in NRMSE. In the case of accuracy, however, the individual values for categorical attributes are compared directly. Thus, a larger spread has a negative effect on the MVI quality.

\paragraph{Conclusion}
The objective of the analysis was to answer the research question "\textit{How do different constraints and input data influence the quality of MVI?}" For this purpose, we investigated which MVI methods prove to be the most suitable. Like Jeong et al.~\cite{Jeong2022} we come to the conclusion that there is no generally best method. The KNNI and MICE methods showed good results, comparable to those of neural networks. These involve more computational effort, but showed less variability in the results than KNNI and MICE. As expected, mean and median imputation performed the worst. Overall, however, the choice of the MVI method seems to depend strongly on the respective data basis.
\pagebreak

With regard to the influencing factors, we made the following findings: 
\begin{itemize}
    \item[\textit{Missing patterns}:] Multivariate and monotone patterns led to slightly worse results in comparison to univariate and general patterns. The reason for this is that these patterns have more dependencies, which complicates the MVI.
    \item[\textit{Missing rate}:] As expected, an increasing missing rate resulted in worse quality of the MVI.
    \item[\textit{Missing mechanism}:] MNAR showed marginally worse results than MAR and MCAR, as anticipated.
\end{itemize}
Overall, however, only the missing rate proved to be decisive. The other factors were not dominant. Further study cases are necessary here.

\subsection{Machine Learning}
As described in Section~\ref{sec:research-methodology}, the analysis includes a classification and pattern recognition for the respective target attribute. In addition to the influencing variables \textit{missing pattern},\textit{ missing rate}, \textit{missing mechanism} and \textit{MVI method} already considered for the MVI, the MVI strategy is also considered here, with a differentiation between \textit{imputed data}, \textit{zerofill} and \textit{delete} approaches. A comparison is always made with the original data. In this way, the influence of the MVI on the results of the machine learning models can be assessed.

Neural networks are denoted with \textit{mlp\-<\_neurons\-\_per\_layer>}, SOM models are denoted with \textit{som<\_neurons\-\_output-layer>}

\begin{figure}[ht]
	\begin{center}
		\includegraphics[width=0.8\linewidth]{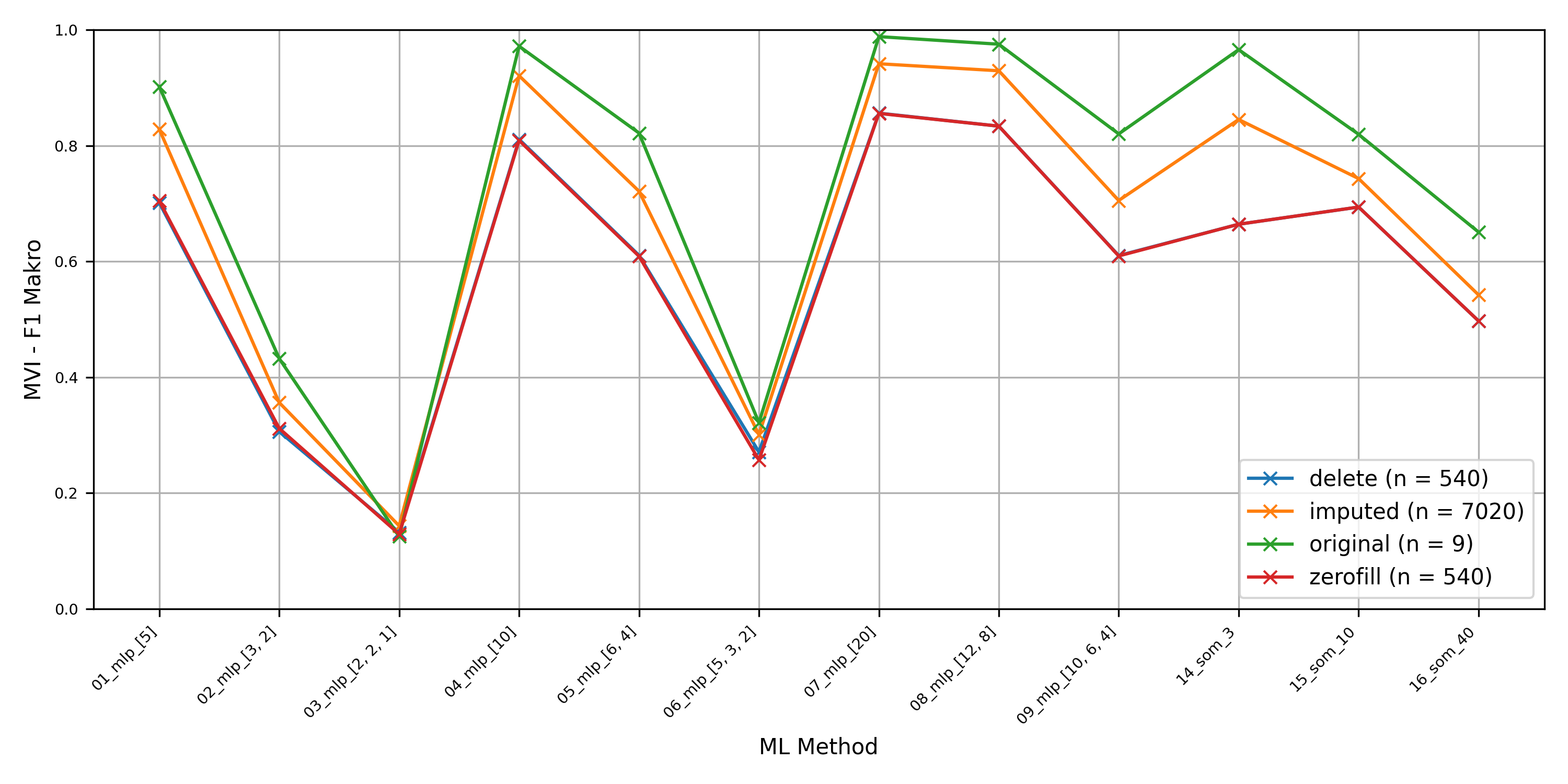}
		\caption{Average $F_1$-makro of the PersonPseudo Data depending on the \textit{MVI strategy} and in comparison to the original data.}
		\label{fig:f1makro-mvi-strategy}
	\end{center}
\end{figure}

If we look at the impact of the MVI strategy on the result of the models, there are distinct gradations. This can be seen in Figure~\ref{fig:f1makro-mvi-strategy}.
The best results were achieved with the original data. Imputed data led to slightly worse results. Nevertheless, an improvement of the F1 makro by approx. 10\% could be obtained for all machine learning models compared to omitting the MVI by zerofill or delete. Zerofill and delete show almost identical results.

The analyses also show that the better the MVI results, the better the performance of the machine learning models. This suggests that the quality of the MVI method is directly related to the subsequent prediction performance.

\begin{figure}[ht]
	\begin{center}
		\includegraphics[width=0.8\linewidth]{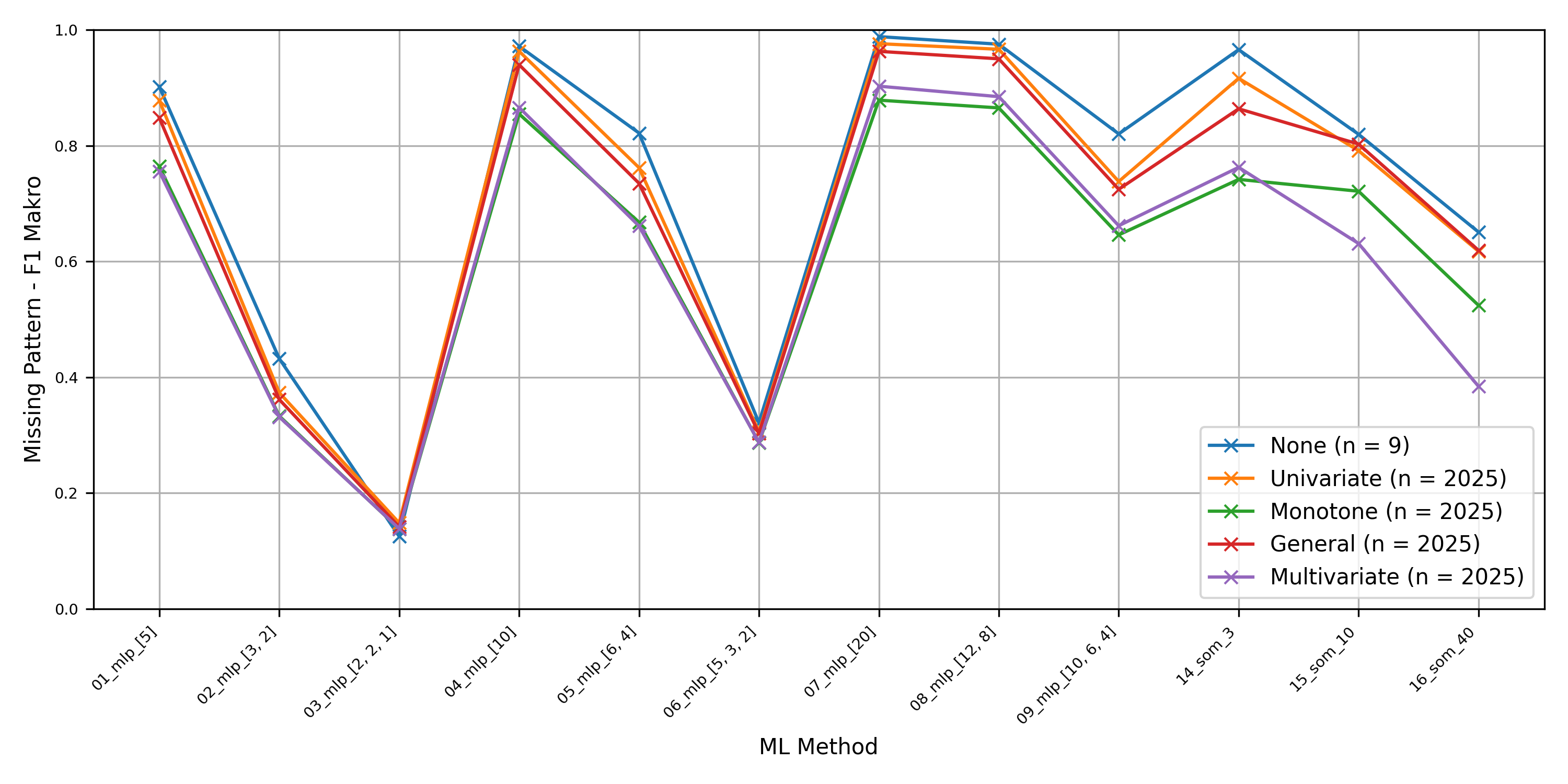}
		\caption{Average $F_1$-makro of the PersonPseudo Data depending on the \textit{missing pattern}.}
		\label{fig:f1makro-missing-pattern}
	\end{center}   
\end{figure}

This also becomes apparent when considering the other influencing variables missing pattern, missing rate and missing mechanism. As with the MVI, the results for missing patterns involving the amputation of single attribute values are superior to those of a joint amputation across multiple attribute values. This is shown by Figure~\ref{fig:f1makro-missing-pattern}.

\begin{figure}[ht]
	\begin{center}
		\includegraphics[width=0.8\linewidth]{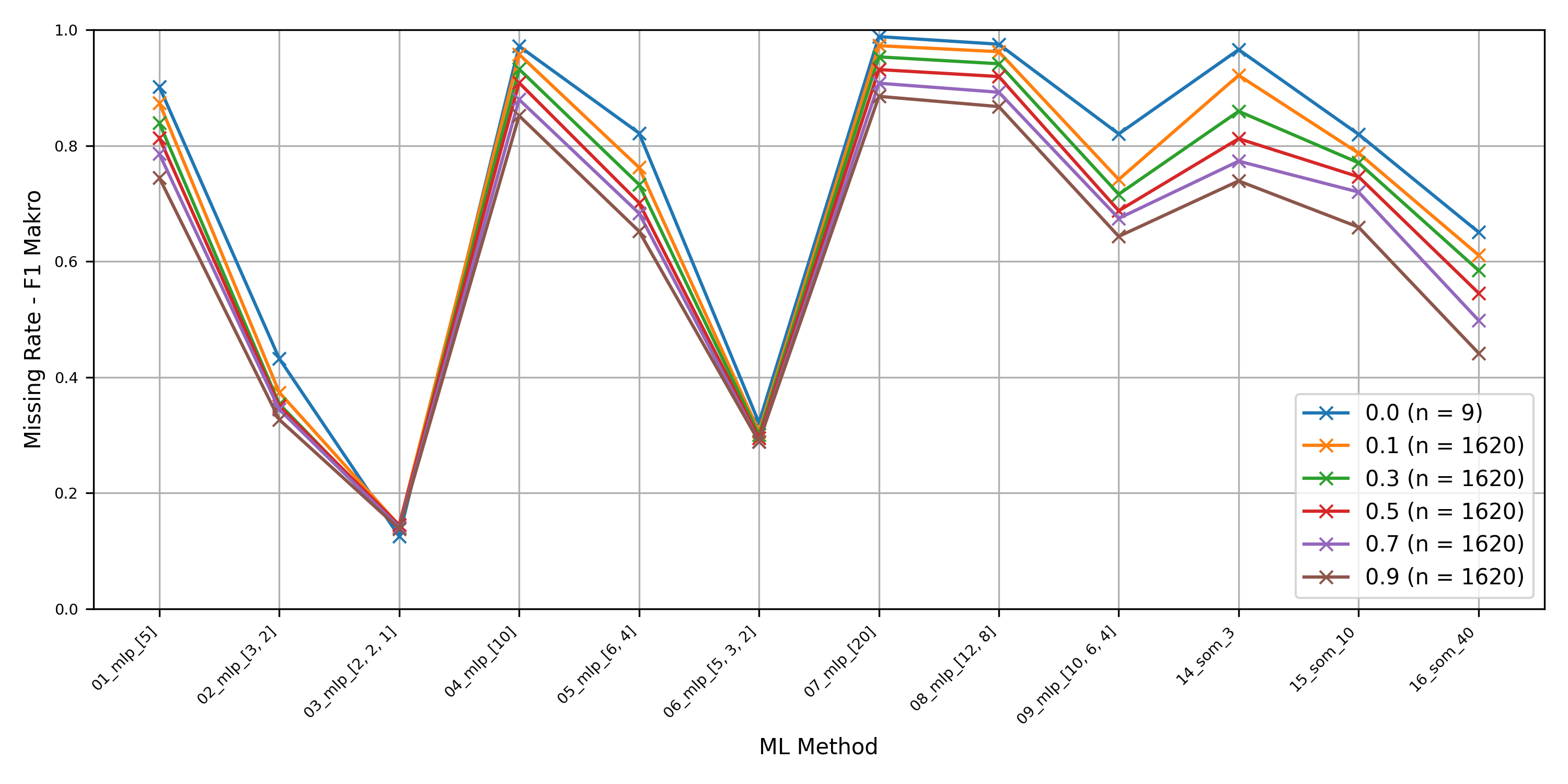}
		\caption{Average $F_1$-makro of the PersonPseudo Data depending on the \textit{missing rate}.}
		\label{fig:f1makro-missing-rate}
	\end{center}
\end{figure}

Likewise, the average F1 makro value decreases as the missing rate increases, as can be seen in Figure~\ref{fig:f1makro-missing-rate}.

\begin{figure}[ht]
	\begin{center}
		\includegraphics[width=0.8\linewidth]{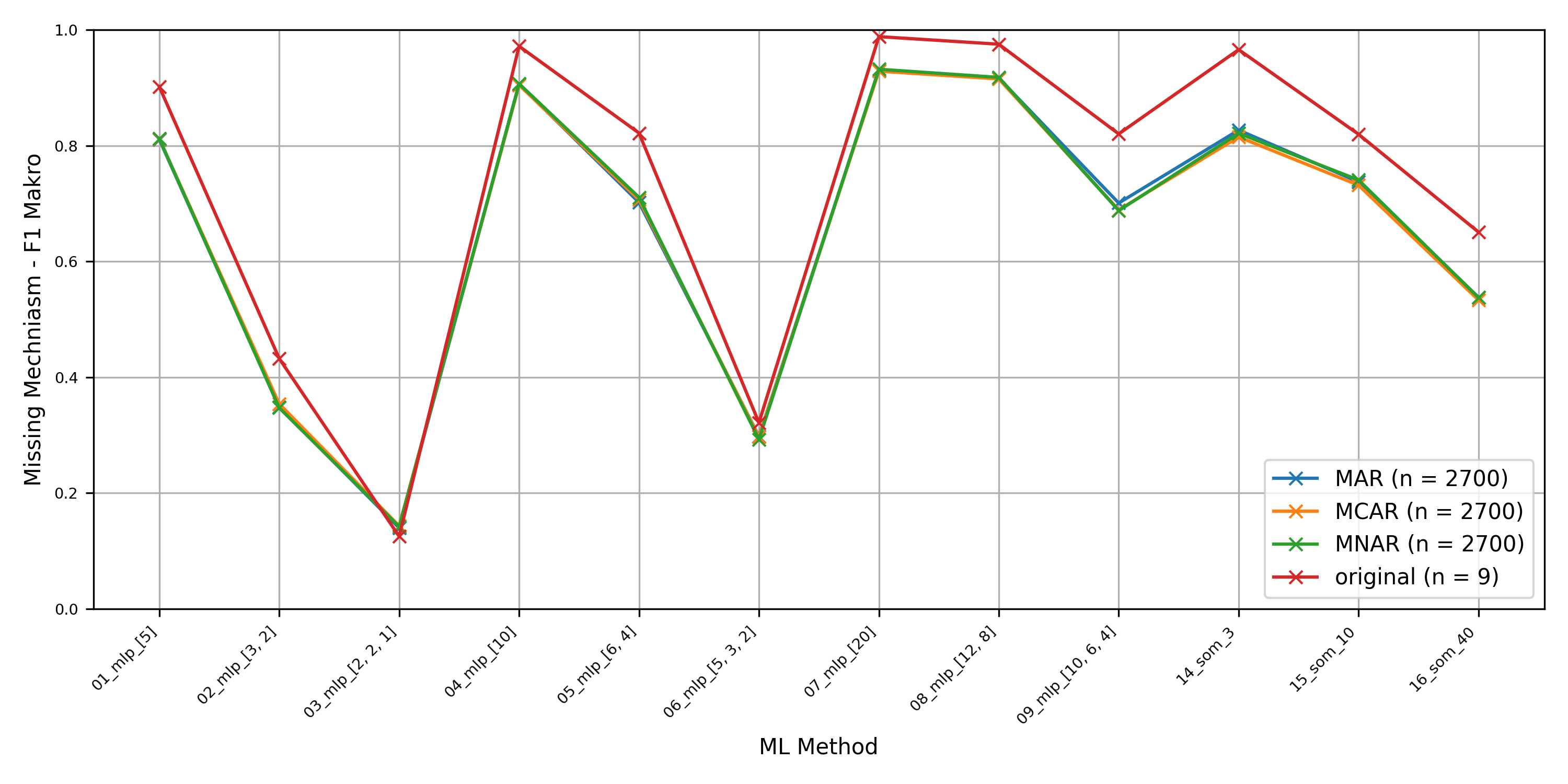}
		\caption{Average $F_1$-makro of the PersonPseudo Data depending on the \textit{missing mechanism}.}
		\label{fig:f1makro-missing-mechanism}
	\end{center}
\end{figure}

In the case of missing mechanism, shown in Figure~\ref{fig:f1makro-missing-mechanism}, only minor differences are evident. 
In all cases, the original data again provide the best F1 makro values.

\paragraph{Conclusion}
The purpose of this section was to answer the research question: "\textit{What is the relation between MVI and the results of downstream machine learning procedures?}". It has been shown that better performance of the MVI also indicates a trend towards better predictive performance of the machine learning models. This leads to the second research question in this context: "\textit{Can MVI also be neglected in the context of machine learning procedures?}". Here it could be shown that the original data led to the best results in classification and pattern recognition. This was followed by the imputed data. As described, better MVI also led to better model performance. The implicit strategies with delete and zerofill produced the worst results. This leads to the conclusion that the MVI plays a very important role. It should not be neglected in the context of machine learning models.

\section{Visualization}
\label{sec:visual}
The results show that MVI is a complex subject that is influenced by different factors. Further work is needed to analyze the impact of different missing rates, missing mechanisms and missing patterns in more detail. A possible first step is an initial analysis by means of visualizations. These can help to better understand the nature of the data and missing values.

In the following, exemplary visualizations are shown with which the individual characteristics of missing data can be better identified and thus analyzed. The R package VIM~\cite{R-VIM} provides various methods for this purpose. All figures in this section were created using this package. The missing values are always shown in red. If the paper is displayed in black and white, the missing values can be recognized by the grayscale contrast.

\begin{figure}[bth]
    \centering
  \subfloat[Aggregation plot]{
     \label{fig:agreggationplot}
     \includegraphics[width=0.7\linewidth]{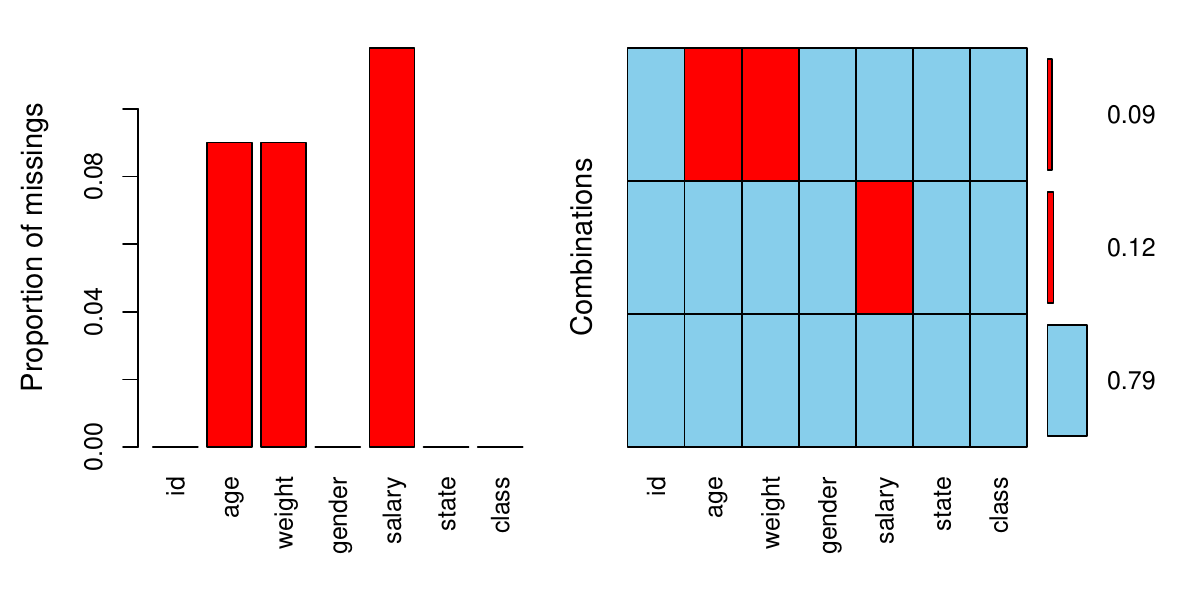}
   } \quad
   \subfloat[Parallel boxplot, distribution of \textit{id}]{
     \label{fig:boxplot-id}
     \includegraphics[width=0.55\linewidth]{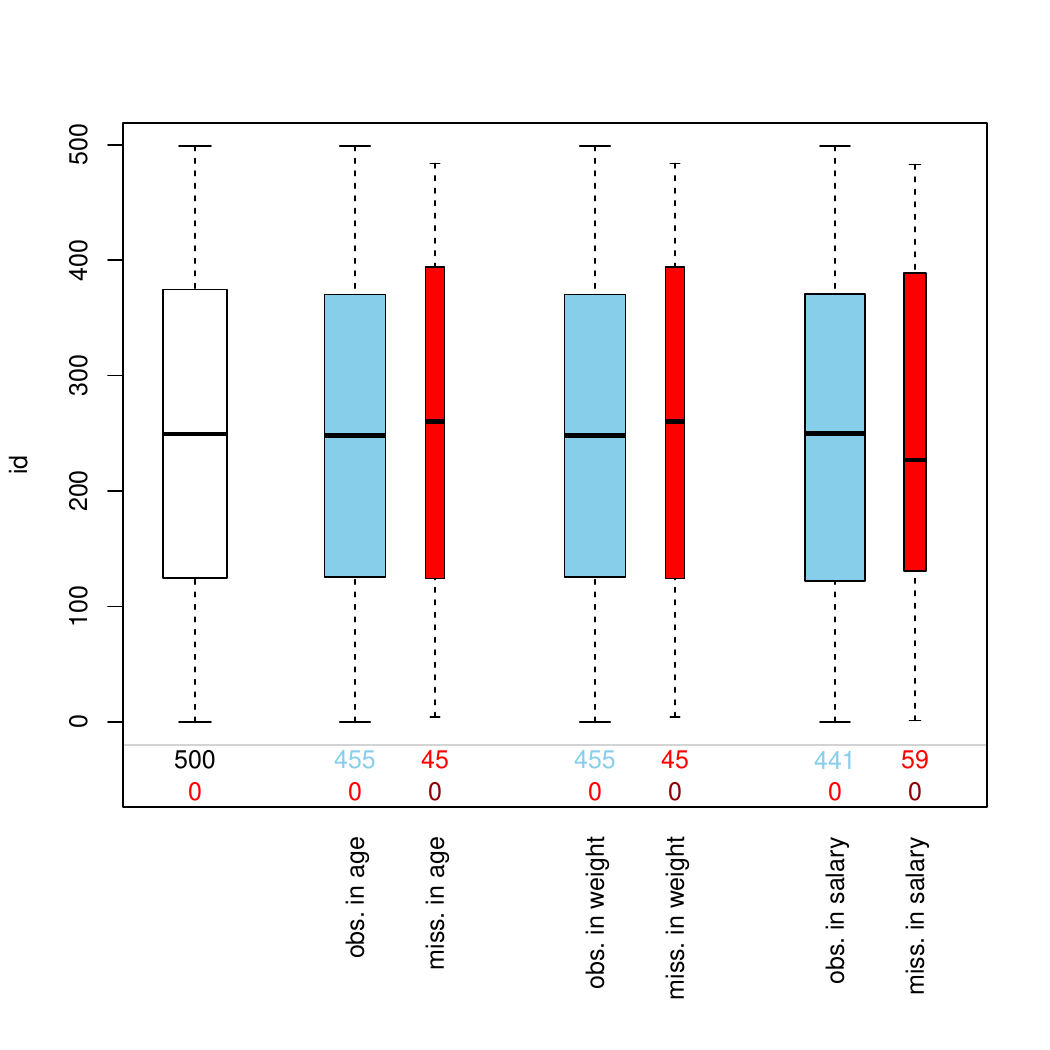}
   }
   	\caption{Identifying the \textit{missing rate} in exemplary extract of PersonPseudo Data. Missing values are colored in red (dark for black and white).}
 	\label{fig:plots-missing-rate}
\end{figure}

\paragraph{Missing rate}
An aggregation plot can be used for the missing rate. This is shown in Figure~\ref{fig:agreggationplot}. 
On the left, the missing rates per variable are shown in a bar plot. On the right, it can be seen in which combinations the missing values occur and how often. This is also helpful for analyzing the missing patterns, as will be described subsequently. It can be clearly seen that the largest missing rate occurs in attribute \textit{salary}. In addition, there are also missing values in \textit{age} and \textit{weight} which have the same missing rate.

This information can also be read from a parallel boxplot. This is shown in Figure~\ref{fig:boxplot-id}.
In this example, the distribution \textit{id} is shown in addition to information about the incomplete attributes. It is also clear to see that \textit{salary} has the highest missing rate, followed by \textit{age} and \textit{weight} with the same rate.

\paragraph{Missing pattern}
Missing patterns can also be detected with the aggregation plot. Figure~\ref{fig:agreggationplot} clearly shows on the right-hand side that there is a multivariate pattern (\textit{age} and \textit{weight}) and a univariate pattern (\textit{salary}).

\begin{figure}[bth]
    \centering
  \subfloat[Matrixplot, sorted by \textit{id}]{
     \label{fig:matrixplot}
     \includegraphics[width=0.45\linewidth]{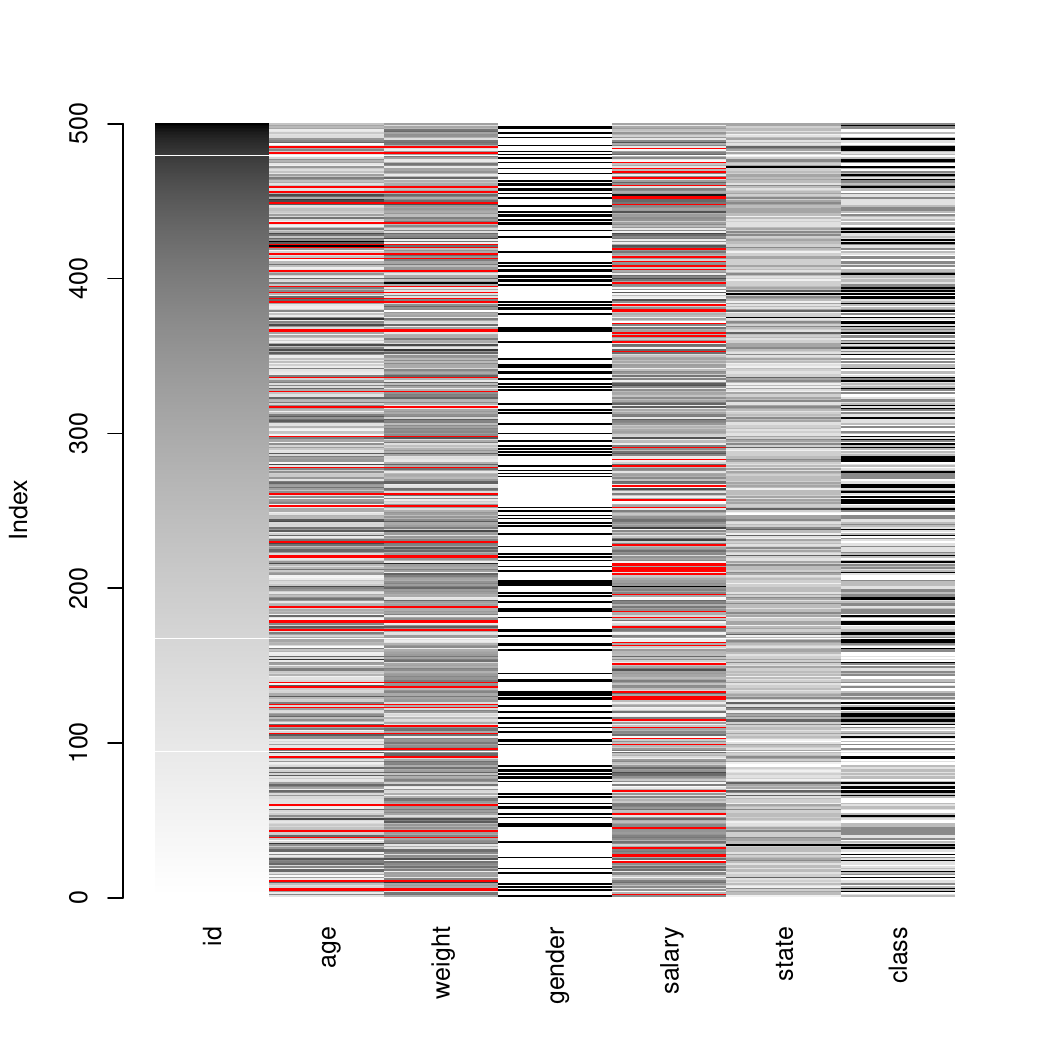}
   } \quad
   \subfloat[Matrixplot, sorted by \textit{age}]{
     \label{fig:matrixplot-age}
     \includegraphics[width=0.45\linewidth]{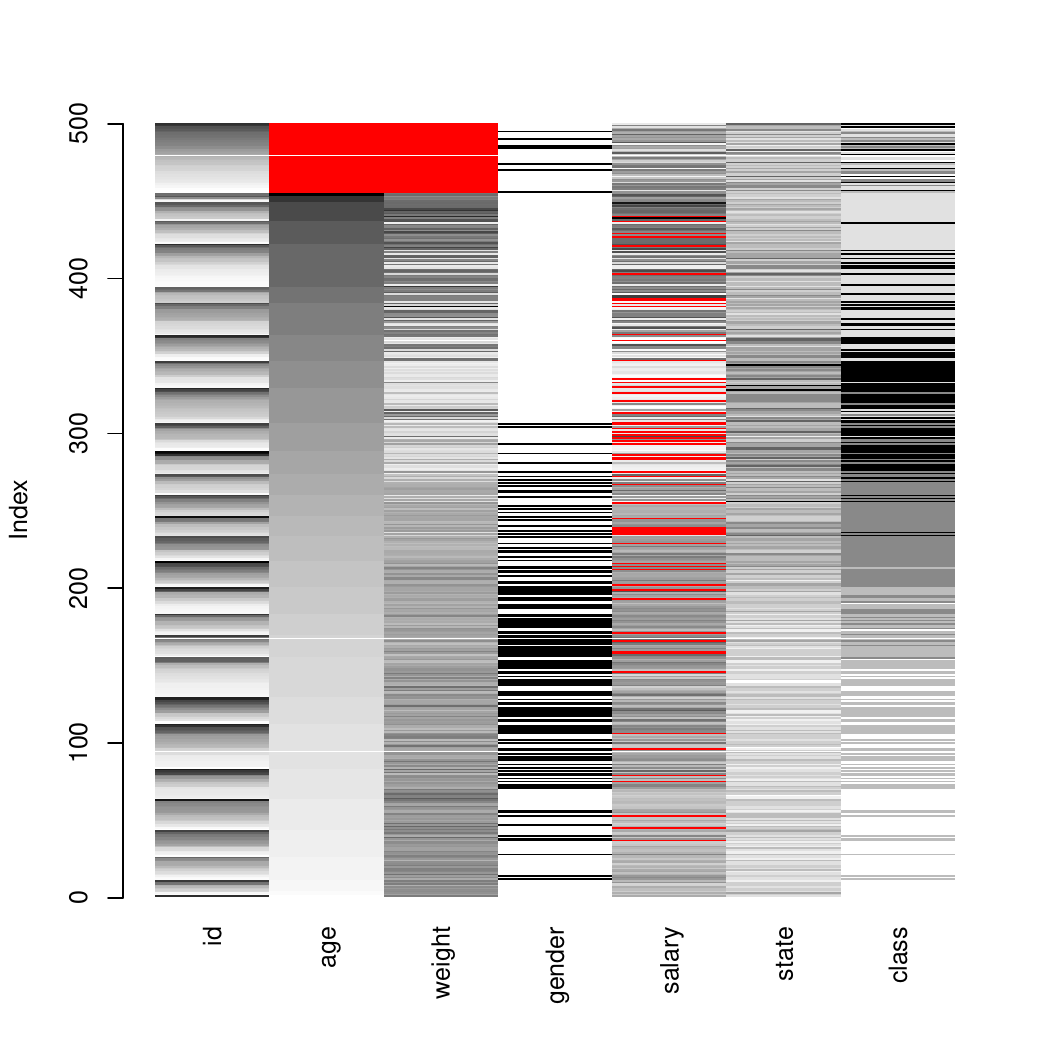}
   } \quad
   \subfloat[Parallel boxplot, distribution of \textit{age}]{
     \label{fig:boxplot-age}
     \includegraphics[width=0.55\linewidth]{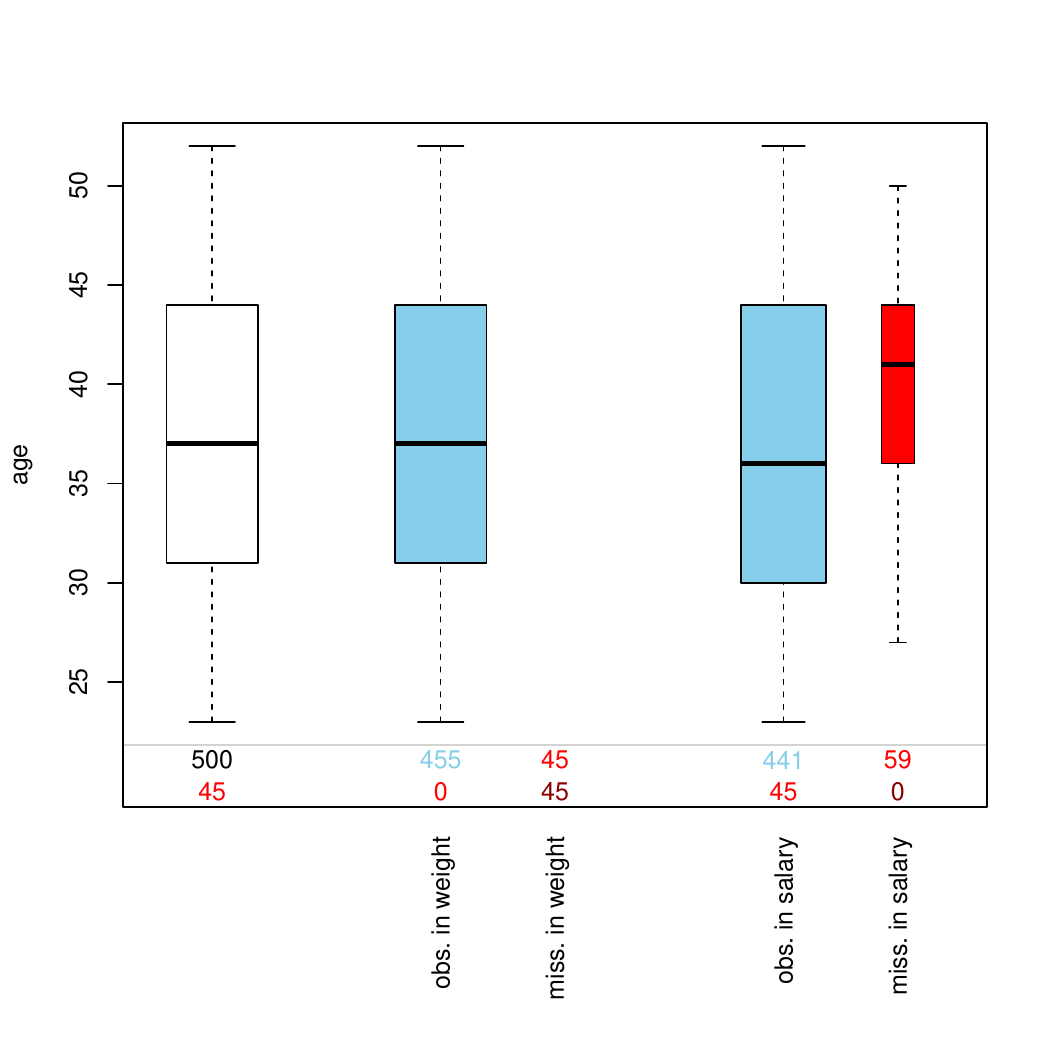}
   }
   	\caption{Identifying the \textit{missing pattern} in exemplary extract of PersonPseudo Data. Missing values are colored in red (dark for black and white).}
 	\label{fig:plots-missing-pattern}
\end{figure}

Another way of identifying missing patterns is the matrix plot, shown in Figure~\ref{fig:matrixplot}.
The values are sorted by the first attribute (\textit{id}). The red (dark for black and white) lines represent missing values. It can already be roughly seen that missing values in \textit{age} and \textit{weight} occur at the same place. It becomes more obvious if the matrix plot is sorted by \textit{age}. This is shown in Figure~\ref{fig:matrixplot-age}. Here, the multivariate pattern of age and weight is very clearly visible. Alternatively, it could be sorted by weight.

A conceivable procedure would be an initial analysis using the aggregation plot, followed by a detailed analysis based on the sorted matrix plot.

In addition, the parallel boxplot can be used here as well. While the multivariate pattern is not yet visible in Figure~\ref{fig:boxplot-id}, it becomes evident when the distributions are plotted as a dependence of \textit{age}, as shown in Figure~\ref{fig:boxplot-age}.
Here it is obvious that if values are missing in \textit{weight}, they are also always missing in \textit{age}. The distributions of \textit{weight} can also be depicted in the same way.

\paragraph{Missing mechanism}
In the case of missing mechanisms, only MCAR and MAR can be identified visually. With MNAR, as described in Section~\ref{sec:characteristics}, missing values depend on the value expression of the same attribute. A visualization is not possible because the values are not present. Accordingly, only MCAR and MAR can be detected. This can be done by means of the matrix plot. However, a MAR mechanism can only be recognized properly if the plot is sorted accordingly.

\begin{figure}[bth]
    \centering
  \subfloat[Matrixplot, sorted by \textit{class}]{
     \label{fig:matrixplot-class}
     \includegraphics[width=0.45\linewidth]{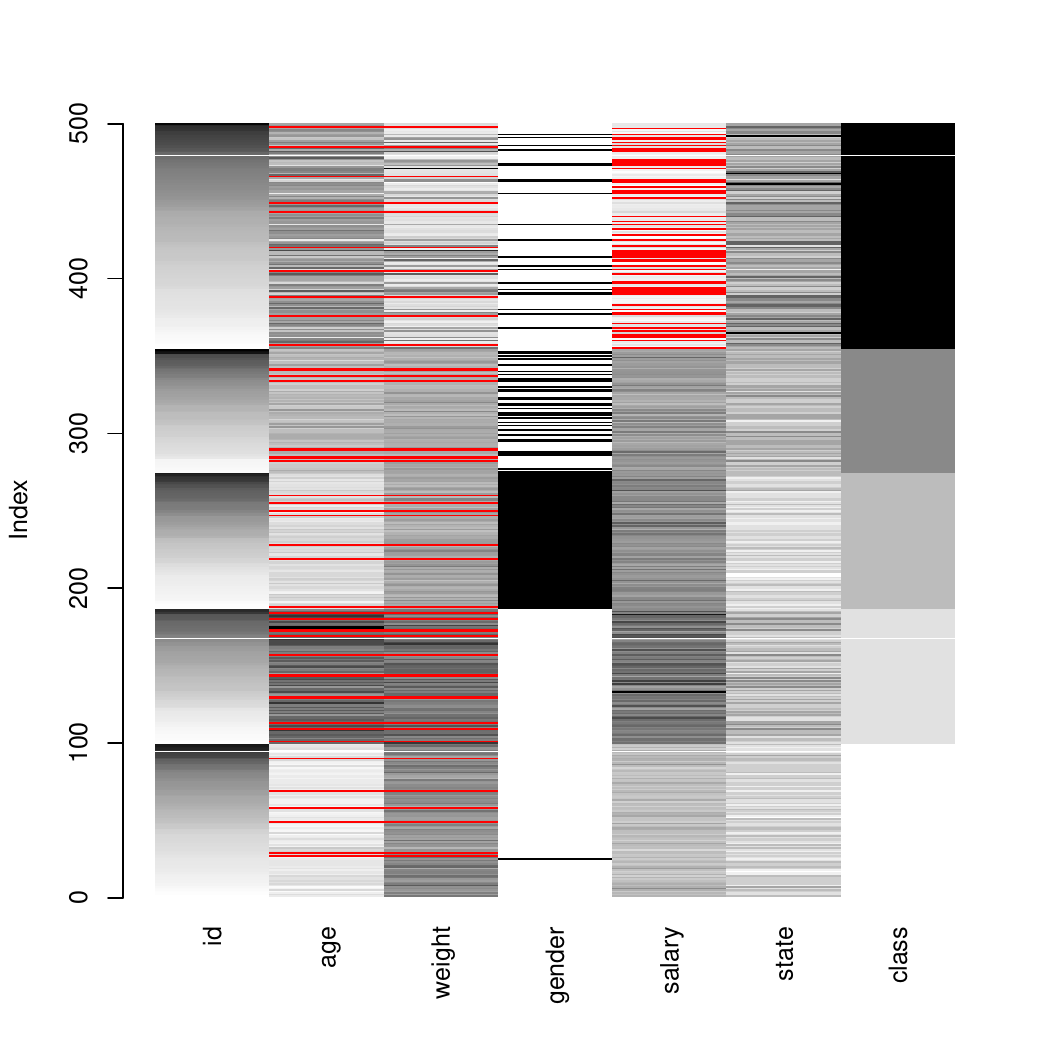}
   } \quad
   \subfloat[Parallel boxplot, distribution of \textit{class}]{
     \label{fig:boxplot-class}
     \includegraphics[width=0.45\linewidth]{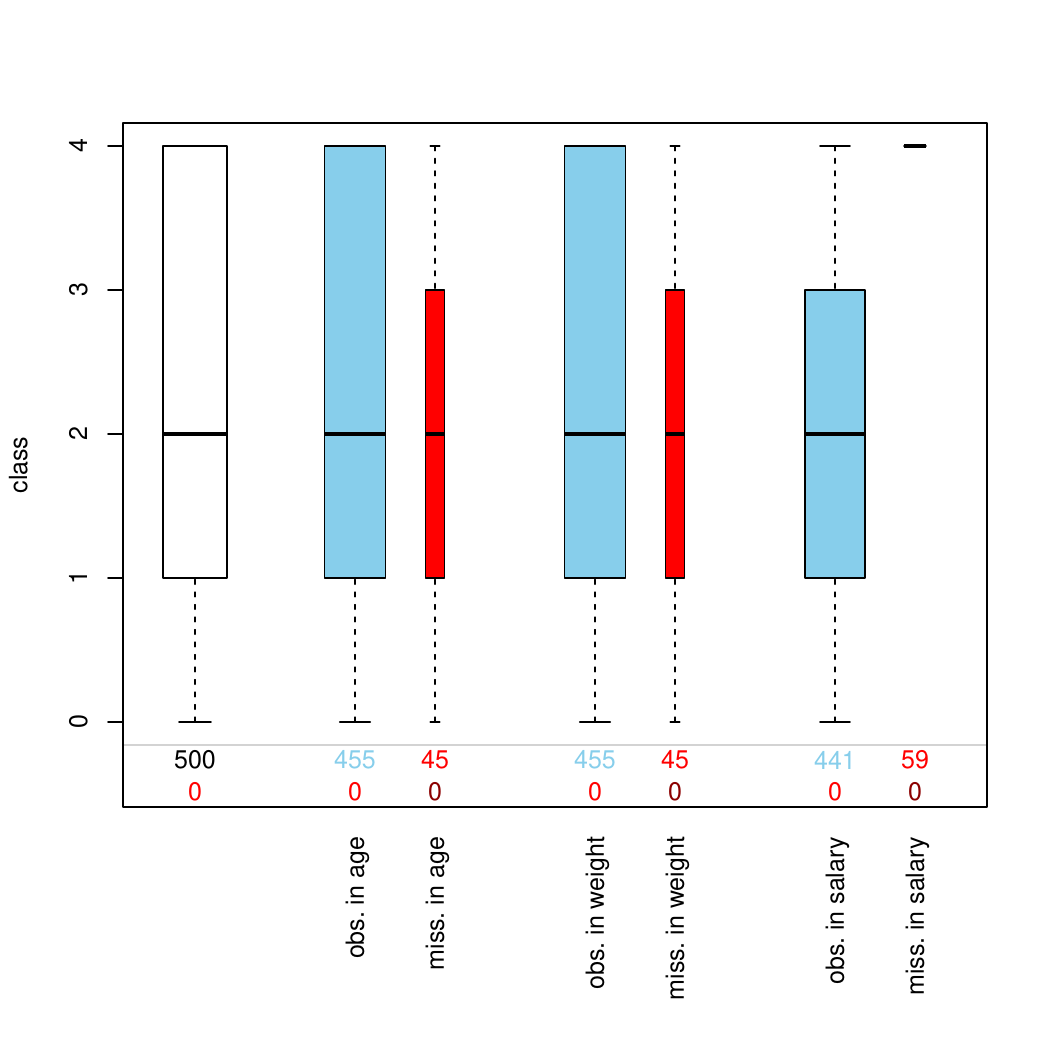}
   }
   	\caption{Identifying the \textit{missing mechanism} in exemplary extract of PersonPseudo Data. Missing values are colored in red (dark for black and white).}
 	\label{fig:plots-missing-mechanism}
\end{figure}

In Figure~\ref{fig:matrixplot}, the MAR mechanism is not yet clearly visible. However, if it is sorted by \textit{class}, as shown in Figure~\ref{fig:matrixplot-class}, it can be easily seen that missing values in \textit{salary} only occur when the values in \textit{class} are high. Alternatively, it can also be sorted by \textit{salary}.

The matrix plot is therefore only suitable to a limited extent for identifying the MAR mechanism. However, if a corresponding assumption is made, it can be checked effectively using the matrix plot.

The situation is similar for the parallel boxplot. In Figure~\ref{fig:boxplot-id}, the MAR mechanism is not yet visible. However, it becomes very apparent if one considers the distributions as a dependence of \textit{class}. This can be seen in Figure~\ref{fig:boxplot-class}.
The distribution of the values of \textit{class} in which \textit{salary} is missing (boxplot on the far right) clearly deviates from the other distributions.

\paragraph{Conclusion}
The aggregation plot provides a good overview of missing rate and missing pattern. It is a useful entry point into the analysis. After a first evaluation, matrix plot and parallel boxplot can be used for detailed investigations. Aggregation plot and matrix plot are suitable for numerical as well as ordinal scaled and binary data. In contrast to aggregation plot, however, matrix plot cannot be used for nominal data. Parallel boxplots are only applicable for numerical data.

This section is intended to give a first impression of how an analysis of characteristics of incomplete data can be supported by visualizations. In Templ et al.~\cite{Templ2012}, the plots are presented in more detail and further plots are included, such as scatter plot and parallel coordinate plot. However, those are often not suitable for larger amounts of data, as the plots quickly become unwieldy. The authors of this paper also use the VIM package~\cite{R-VIM} to create their visualizations.

\section{Summary and Outlook}
\label{sec:summary}
In this paper, we presented the MVIAnalyzer, a generic approach for a holistic analysis of MVI. It considers the overall context of data, influences of characteristics of incomplete data and the subsequent evaluation of data quality and ML predictions. This makes it possible to perform independent analyses.

In our current work, we used the MVIAnalyzer to investigate the influences of different constraints on the quality of MVI and downstream ML models. For this purpose, we conducted an extensive analysis of different characteristics of incomplete data. These include, among others, the following, often neglected, aspects: \textit{Missing rates over 50\%}, \textit{various missing patterns}, \textit{besides MCAR the consideration of MAR and MNAR}.

It could be shown that the quality of the MVI also affects the quality of the prediction performance of the models. Hence, it can be concluded that MVI should not be neglected in the context of machine learning methods. Regarding the influences of the constraints, it can be seen that the explicit MVI consistently achieved better results than the implicit strategy. Regarding the characteristics of incomplete data, it was shown that especially an increasing missing rate has a negative impact on the outcome. However, the results also show the complexity resulting from the different characteristics of incomplete data. Further investigation of individual aspects is therefore necessary.

The MVIAnalyzer presented serves as a good basis for this. It provides a generic approach for comparison of MVI methods. Various characteristics of incomplete data can be taken into account and systematized as thoroughly as possible by means of a missing value simulation. The framework can also be used by other researchers with their own data. The configuration file provides many different options for extensive analyses.

In addition, the analysis can be specifically aided by visualizations. In this paper, we have presented different visualizations and their use for different data and characteristics of incomplete data.

\section*{Acknowledgement}
Thanks to André Conrad for his help in providing the code and the docker and git tips!

\bibliographystyle{plainurl}  
\bibliography{bibliography}  

\end{document}